\documentclass[12 pt,oneside,onecolumn,a4paper]{article}
\usepackage{epsfig,graphicx}
\usepackage{cite}
\usepackage{indentfirst}
\usepackage{amsfonts,amssymb,latexsym,amsmath,enumerate,amsthm}
\usepackage{indentfirst,cite}
\usepackage{bm}
\def\rot{{\rm rot}}

mm \leftmargin 10pt \rightmargin 5pt \hoffset= -10mm

\title{On the theory of polarization radiation generated\\ in the media with sharp boundaries}

\author{D.V. Karlovets\thanks{e-mail: d.karlovets@gmail.com. This manuscript deals with problems some of which were treated before in the preprint arXiv:0908.2336v2. 
However, results obtained there, strictly speaking, are not applicable for transparent media ($\varepsilon^{\prime \prime} = 0$) because the multiple reflections of radiation inside the medium were not taken into account. In the present paper, this defect has been corrected.}}

\date{\small{Tomsk polytechnic university, \\Lenina 30, 634050 Tomsk, Russian Federation}}
\begin{document}
\maketitle

\begin{abstract}
Polarization radiation arising when a charged particle moves uniformly in vacuum nearby the media possessing a finite permittivity $\varepsilon (\omega) = \varepsilon^{\prime} + i \varepsilon^{\prime \prime}$ and sharp boundaries is considered. The method is developed in which polarization radiation is represented as a field of the current density induced in matter by the field of the moving charge. Solution is found for a problem of radiation arising when the particle moves along the axis of the vacuum cylindrical channel inside a thin screen of a finite radius and a finite permittivity. Depending on parameters of the problem, the solution obtained describes different types of polarization radiation: Cherenkov radiation, transition radiation and diffraction radiation. In particular, when the channel radius approaches zero and external radius of the screen tends to infinity the expression found for radiated energy completely coincides with the one derived by Pafomov for transition radiation generated in a slab. In another special case of ideal conductivity, the result obtained coincides with the one for diffraction radiation generated by the particle in the round hole in the infinitely thin screen. Solution is found for a problem of radiation generated when the charge moves nearby a rectangular screen possessing a finite permittivity. The expression derived describes diffraction and Cherenkov mechanisms of radiation and takes into account the possible multiple reflections of radiation inside the screen. Solution is also found for a problem of polarization radiation generated when the particle moves nearby a thin grating consisting of a finite number of rectangular strips possessing a finite permittivity and separated with vacuum gaps (Smith-Purcell radiation). In the special case of an ideally conducting grating, the formula derived for radiated energy coincides with the one of the well-known surface current model.
\end{abstract}

\section{\large{Introduction}}
\label{Sect0}

The field of a charged particle moving uniformly and rectilinearly in a medium may lead to dynamic polarization of atomic shells and emission of radiation known as polarization radiation in the microscopic
theory \cite{A}. In the m\textit{a}croscopic theory, the motion of a particle through a medium (or near a medium) with sharp boundaries leads to various types of polarization radiation: Cherenkov radiation (in the absence of boundaries), transition radiation (the particle intersects the interface between two media), diffraction radiation (the particles moves near the interface), etc. The number of exact solutions to the problem of polarization radiation for media with sharp boundaries is not large. The most important of the exactly solvable problems are the calculations of transition radiation in a slab of a finite permittivity \cite{B} and diffraction radiation from a perfectly conducting semi-plane \cite{J, PLA} and wedge \cite{VDB, G}. The solution of such problems in more complex geometries is complicated by difficulties in the formulation of the boundary conditions. It is well known, for example, that rigorous solution of the problem of radiation from a charge moving near a perfectly conducting grating (so-called Smith-Purcell radiation) is expressed by infinite series and leads to a system of an infinitely large number of equations \cite{Sh, Kube}. However, surfaces with intricate profiles are most interesting for applications. For this reason, the development of approximate methods in the problems of polarization radiation is topical.

In this paper, an approach is developed in which polarization radiation is represented as the field of the current induced in a substance by the field of a point charge moving uniformly in vacuum. The advantage of this method lies in the possibility of determining the characteristics of various types of polarization radiation (including the case when several types of radiation are generated simultaneously) for a wide range of surfaces taking into account actual dielectric properties of the substance and frequency dispersion. It will be shown that, in some special cases, the results obtained using the method developed here are transformed into familiar expressions for Cherenkov radiation, transition radiation, and diffraction radiation (as well as parametric X-ray radiation; see \cite{Mono}).

The paper is organized as follows. In Section 2, the method of induced currents is described and the radiation field generated during the motion of a charge along the axis of a cylindrical vacuum channel in a screen with finite permittivity $\varepsilon (\omega) = \varepsilon^{\prime} + i \varepsilon^{\prime \prime}$, thickness, and radius is described. Then, the emergence of radiation from the substance into vacuum is considered, and the energy emitted into vacuum in the ``forward'' and ``backward'' directions is determined using the well-known reciprocity theorem. In Section 3, the solution to the problem of radiation generated when a particle moves near a rectangular screen of finite sizes and permittivity is obtained. In contrast to the known solution, possible multiple re-reflections of the radiation field in the screen are taken into account. In Section 4, the solution to the problem of radiation generated by a particle moving in the vicinity of a thin grating consisting of rectangular strips and possessing a finite permittivity is obtained. The results, including the limits of  their applicability, are discussed in Section 5.

\section{\large{Polarization Radiation Generated in a Cylindrical Channel in a Screen of a Finite Thickness and Radius}}
\label{Sect1}

\subsection{Radiation inside a Screen with a Finite Permittivity}
\label{Sect1.1}

Let us consider a homogeneous isotropic non-magnetic medium with a complex conductivity and a frequency dispersion:
\begin{eqnarray}
\displaystyle && \sigma (\omega) = \frac{i \omega}{4 \pi}(1 - \varepsilon (\omega)). \label{Eq1}
\end{eqnarray}  
A charged particle possessing an energy $\gamma = E/mc^2 = 1/\sqrt{1-\beta^2}$, and moving at a constant velocity $v = \beta c$ in the substance induces in it a polarization currents of density:
\begin{equation}
\displaystyle \bold {j}_{pol} = \sigma (\bold E^0 + \bold E^{pol} (\bold {j}_{pol})), \label{Eq2}
\end{equation}
where $\bold E^0 \equiv \bold E^0 (\bold r, \omega), \bold E^{pol} \equiv \bold E^{pol} (\bold r, \omega)$  are the Fourier transforms of the field of the particle in vacuum and of the field of currents induced in the substance.

To determine the field of polarization radiation propagating in the substance in the form of transverse waves, we must solve the ``vacuum'' Maxwell equations with current density $\bold {j}_{pol}$ on the right-hand side. However, Eq.(\ref{Eq2}) for the current density is an integral equation. The integral term can be disregarded if the conductivity of the medium is low, $|\varepsilon -1| \ll 1$, which
makes it possible to obtain the required solutions in the X-ray part of the spectrum \cite{D, Mono, T-PRE, T-PLA, Sysh}. In the general case, when parameter $|\varepsilon -1|$ is not small, the field acting on each atom and molecule does not coincide with vacuum field $\bold E^{0}$ any longer. However, in this case also we can proceed in the same way as in deriving the macroscopic Maxwell equations. Transferring the integral term $\sigma \bold E^{pol}$ to the left-hand side of these equations, we obtain the following equation for the magnetic field:
\begin{eqnarray}
\displaystyle \Big (\Delta + \varepsilon (\omega) \frac{\omega^2}{c^2}\Big ) \bold H^{pol} (\bold r, \omega ) = - \frac{4 \pi}{c}
\sigma (\omega) \rot \bold E^0 (\bold r, \omega ). \label{Eq3}
\end{eqnarray}
The solution to this equation in the wave zone gives the field of polarization radiation emitted by atoms and molecules of the substance as a result of so-called distant collisions, in which the energy lost by a particle is negligibly low as compared to its total energy. It will be shown below that in an unbounded medium this solution coincides with the field of Cherenkov radiation.

If the polarization currents are induced in a limited volume of the substance (i.e., the medium has boundaries), the integration in the solution of (\ref{Eq3}) is performed only over the domain $V_T$ occupied by the currents:
\begin{eqnarray}
\displaystyle \bold H^{pol} (\bold r, \omega ) = \rot \frac{1}{c} \int \limits_{V_T} \bold \sigma (\omega) \bold E^0 (\bold {r}^{\prime}, \omega) 
\frac{e^{i \sqrt{\varepsilon (\omega )} \omega |\bold r - {\bold r}^{\prime}|/c}}{|\bold r - {\bold r}^{\prime}|} d^3 r^{\prime}. \label{Eq4}
\end{eqnarray}
In a bounded volume of the substance, the waves reflected from the boundaries, which cannot be described by expression (\ref{Eq4}), also exist apart from
polarization radiation waves. However, such waves can be taken into account at a later stage when the emergence of radiation from the medium into vacuum
will be considered. Evaluating the curl in Eq.(\ref{Eq4}) we obtain 
\begin{eqnarray}
\displaystyle && \bold H^{pol}(\bold r, \omega ) =  \frac{1}{c} \int \limits_{V_T} \frac{\bold r - \bold r^{\prime}}{|\bold r - \bold r^{\prime}|} \times \sigma (\omega) \bold E^0(\bold r^{\prime}, \omega ) \Big (i \sqrt{\varepsilon} \frac{\omega}{c} - \frac{1}{|\bold r - \bold r^{\prime}|} \Big ) \frac{e^{i \sqrt{\varepsilon} \omega |\bold r - {\bold r}^{\prime}|/c}}{|\bold r - {\bold r}^{\prime}|} d^3 r^{\prime} \label{Eq5}
\end{eqnarray}
(here and below dependence of $\varepsilon$ on frequency is implied). 

It can be seen from this expression that the wave zone of radiation is defined by the following inequalities:
\begin{eqnarray}
\displaystyle && r \gg \lambda/\sqrt{\varepsilon (\omega)}, \ r \gg r^{\prime}_{eff}, \label{Eq10}
\end{eqnarray}
where $r^{\prime}_{eff}$ determines the effective region in the substance, occupied by currents and making the main contribution to integral (\ref{Eq5}). It should be noted that the
first of the conditions of the wave zone cannot be satisfied near permittivity zeros ($\varepsilon (\omega) \rightarrow 0$). This corresponds to the well-known fact that there are no
transverse waves propagating in the substance to infinity at resonant frequencies \cite{Ryaz-RF}. When conditions (\ref{Eq10}) are satisfied, the expression for the field of induced
currents describes the transverse radiation field in the substance:
\begin{eqnarray}
\displaystyle \bold H^{pol} \approx \bold H^{R} = \sqrt{\varepsilon}\frac{i \omega}{c^2} \frac{e^{i r \sqrt{\varepsilon} \omega/c}}{r} \ \bold e \times \int \limits_{V_T} \sigma (\omega) \bold E^0(\bold r^{\prime}, \omega ) e^{- i \bold k \bold r^{\prime}} d^3 r^{\prime}, \label{Eq11}
\end{eqnarray}
where $\bold k = \sqrt{\varepsilon} \bold {e} \omega/c$ is the wave vector of radiation in the medium. It follows from the Maxwell equations that the electric field strength is defined as
\begin{eqnarray}
\displaystyle \bold E^{pol} \approx \bold E^{R} = - \frac{1}{\sqrt{\varepsilon}} \ \bold e \times \bold H^{R}. \label{Eq11a}
\end{eqnarray}

Let us consider radiation generated during uniform rectilinear motion of a charged particle along the axis of a cylindrical vacuum channel of radius $a$ in a screen of thickness $d$ and outer radius $b$ (see Fig.1a.). To determine the radiation field, we must evaluate the volume integral in expression (\ref{Eq11}), in which the field of the particle is defined by the well-known expression (see, for example, \cite{Topt}):
\begin{eqnarray}
&& \displaystyle \bold{E}^0(\bold {r}, \omega) = \frac {e \omega}{\pi v^2 \gamma}\Big(\frac {{\bm \rho}}{\rho}K_1 \Big[\frac {\omega \rho}{v \gamma}\Big] -
\displaystyle \frac {i}{\gamma}\frac {\bold v}{v}K_0 \Big[\frac{\omega \rho}{v \gamma}\Big]\Big) e^{i\displaystyle \frac {\omega}{v} z}, \label{Eq12}
\end{eqnarray}
where ${\bm \rho} = \{x, y\}$ and $K_0, K_1$ are modified Bessel functions of the second kind. For the magnetic field of radiation, we have
\begin{eqnarray}
\displaystyle && \bold H^{R} (\bold r, \omega ) = \frac {e \omega^2 (\varepsilon - 1)}{4 \pi^2 v^2 c \gamma} \frac{e^{i r \sqrt{\varepsilon} \omega/c}}{r} \bold k \times 
\int \limits_{a}^{b}\rho^{\prime}d\rho^{\prime} \int \limits_{0}^{2 \pi} d\phi^{\prime} \int \limits_{-d}^{0} dz^{\prime}
\Big(\frac{{\bm \rho^{\prime}}}{\rho^{\prime}} K_1 \Big[\frac {\omega \rho^{\prime}}{v \gamma}\Big] - \cr \displaystyle && \qquad \qquad \qquad \qquad \qquad \qquad \qquad \qquad \qquad - \displaystyle \frac {i}{\gamma}\frac {\bold v}{v} K_0 \Big[\frac{\omega \rho^{\prime}}{v \gamma}\Big]\Big) e^{- i \bold k \bold r^{\prime} + i \frac {\omega}{v} z^{\prime}}, \label{Eq13}
\end{eqnarray}
where the unit vector of radiation in the medium $\bold e$ has the form
\begin{eqnarray}
\displaystyle && \bold {e} = \{\sin{\Theta} \cos {\phi}, \sin{\Theta} \sin {\phi}, \cos{\Theta}\}. \label{Eq14}
\end{eqnarray}

It should be emphasized that the polarization radiation field obtained in this way does not contain information on possible reflections of waves from the boundaries of the substance itself or on possible re-reflections of waves in the vacuum channel. Re-reflection of waves in the substance can be taken into account (see Section 3), but allowance for re-reflection of waves in the vacuum channel is impossible in the given approach. If, however, we confine our analysis to the case of not too low energies of the particle ($\beta \precsim 1$), this constraint is substantial only for large thicknesses of the screen because Cherenkov radiation is known to be unable to enter into the vacuum from the medium through the plane parallel to the velocity vector of the particle (total internal reflection; see \cite{Bol-61} and Section 4 below), while diffraction radiation propagates at small angles $\theta \sim \gamma^{-1} \ll 1$ in the relativistic case. To derive a quantitative condition, we assume
that the wave is emitted at angle $\theta = \gamma^{-1}$ at point $z= - d$ and reaches the opposite end of the channel at point $z= 0$ (see Fig.1b.). Then the condition for the absence
of re-reflection of radiation in the channel and, hence, the condition for the applicability of the formulas derived below is the inequality
\begin{eqnarray}
\displaystyle && d \ll d_{max} = 2 a \gamma. \label{Eq15}
\end{eqnarray}
If the radius of the channel is several times larger than the radiation wavelength (otherwise, re-reflections can simply be disregarded), the value of $d_{max}$ is an order of magnitude larger than parameter $\gamma \lambda$. For the terahertz wavelength range and for relativistic energies of the particle, the experimental values of the screen thickness satisfy the conditions for the applicability of the
model because $d_{max}$ in this case amounts to more than
a few tens of centimeters. In the optical range, the model is valid only for high energies ($\gamma \sim 10^{3}$ and higher). Finally, in the X-ray spectral range, reflections can also be disregarded.

Integration in formula (\ref{Eq13}) is carried out using the known relations for Bessel functions \cite{Ryzh}. The result has the form
\begin{eqnarray}
\displaystyle & \displaystyle \bold H^{R} (\bold r, \omega ) = \frac {e}{2 \pi c}\frac{\omega}{c} \sqrt{\varepsilon}(\varepsilon - 1) \frac{e^{i r \sqrt{\varepsilon} \omega/c}}{r} 
\frac{e^{-i d \frac{\omega}{c}(\beta^{-1} - \sqrt{\varepsilon} \cos{\Theta})}- 1}{(1 - \beta \sqrt{\varepsilon} \cos{\Theta})(1 + \varepsilon (\beta \gamma \sin{\Theta})^2)} 
\{\sin {\phi}, - \cos {\phi}, 0\} \cr \displaystyle & \ \Big ( \sin{\Theta} (\gamma^{-1} - \beta \gamma \sqrt{\varepsilon} \cos{\Theta}) \Big (a J_0\Big (a \frac{\omega}{c} \sqrt{\varepsilon}
\sin{\Theta}\Big ) K_1\Big (a \frac{\omega}{v \gamma}\Big ) - b J_0\Big (b \frac{\omega}{c} \sqrt{\varepsilon} \sin{\Theta}\Big ) K_1\Big (b \frac{\omega}{v \gamma}\Big )\Big ) \cr \displaystyle & - (\cos{\Theta} + \beta \sqrt{\varepsilon} \sin^2{\Theta}) \Big (a J_1\Big (a \frac{\omega}{c} \sqrt{\varepsilon}\sin{\Theta}\Big ) K_0\Big (a \frac{\omega}{v \gamma}\Big ) - b J_1\Big (b \frac{\omega}{c} \sqrt{\varepsilon} \sin{\Theta}\Big ) K_0\Big (b \frac{\omega}{v \gamma}\Big )\Big ) \Big ). \label{Eq16}
\end{eqnarray}
In the particular case of a layer of thickness $d$ (for which $a \rightarrow 0, \ b \rightarrow \infty$), we have the following expressions for the Bessel functions depending on variable $a$:
\begin{eqnarray}
\displaystyle && a J_0 K_1 \rightarrow \frac{v \gamma}{\omega}, \ a J_1 K_0 \rightarrow 0. \label{Eq17}
\end{eqnarray}
The functions depending on $b$ vanish in view of the exponential decay of functions $K_0$ and $K_1$ upon an increase in the value of $b$ \cite{Ryzh}. In the case of a transparent medium, the energy emitted in a layer into the unit frequency interval can be determined by the formula
\begin{eqnarray}
\displaystyle \frac{d W}{d \omega} = \int  \frac{c r^2}{\sqrt{\varepsilon}}|\bold H^{R}|^2 d \Omega = \frac{2 e^2 \beta^2}{\pi c}\sqrt{\varepsilon} 
\int \limits_0^\pi \frac{\sin^2{\Big (\frac{d}{2} \frac{\omega}{v} (1 - \beta \sqrt{\varepsilon} \cos{\Theta})\Big )}}{(1 - \beta \sqrt{\varepsilon} \cos{\Theta})^2} 
\times \cr \qquad \qquad \qquad \qquad \qquad \displaystyle \Big [\frac{(\varepsilon - 1) (1 - \beta^2 - \beta \sqrt{\varepsilon} \cos{\Theta})}{1 - \beta^2 + (\beta \sqrt{\varepsilon} \sin{\Theta})^2} \Big ]^2 \sin^3 {\Theta} d \Theta. \label{Eq18}
\end{eqnarray}
In the vicinity of the Cherenkov angle $\cos \Theta = 1/(\beta \sqrt{\varepsilon})$ the terms in the brackets associated with the transition mechanism of radiation become equal to unity. In this
case, formula (\ref{Eq18}) completely coincides with the analogous expression obtained in the theory of Cherenkov radiation in a finite-thickness layer (see, for example, \cite{Z, Gr}).
For a layer with a large thickness, the interference term is transformed into the delta-function:
\begin{eqnarray}
\displaystyle && \frac{\sin^2{\Big (\frac{d}{2} \frac{\omega}{v} (1 - \beta \sqrt{\varepsilon} \cos{\Theta})\Big )}}{(1 - \beta \sqrt{\varepsilon} \cos{\Theta})^2} \rightarrow \frac{d}{2}\frac{\omega}{v} \pi \ \delta \Big (1 - \beta \sqrt{\varepsilon} \cos{\Theta}\Big ), \label{Eq19}
\end{eqnarray}
after which the integration in expression (\ref{Eq18}) leads to the conventional Tamm-Frank formula for Cherenkov radiation in an unbounded transparent medium:
\begin{eqnarray}
\displaystyle && \frac{1}{d}\frac{d W}{d \omega} = \frac{e^2}{c^2}\omega \Big (1 - \frac{1}{\beta^2 \varepsilon}\Big ). \label{Eq20}
\end{eqnarray}

\subsection{Energy Emitted into Vacuum}
\label{Sect1.2}

Polarization radiation field (\ref{Eq16}) determined above has the form of a spherical wave because it was obtained under the assumption that the observer is in the wave zone. However, the field of the wave incident from the absorbing medium on the screen boundary located in the $z=0$ plane (see Fig.1a) naturally differs from (\ref{Eq16}). To solve the problem of refraction of the wave at the interface between the medium and vacuum correctly, we must use the well-known reciprocity principle \cite{L}:
\begin{eqnarray}
\displaystyle && (\bold E^{R(vac)}, \bold d^{(vac)}) = (\bold E^{R(m)}, \bold d^{(m)}), \label{Eq21}
\end{eqnarray}
where $\bold E^{R(vac)}\equiv \bold E^{R(vac)} (\bold r, \omega)$ is the sought radiation field in vacuum, which is produced by a dipole with moment $\bold d$ located in the medium, and
$\bold E^{R(m)} \equiv \bold E^{R(m)} (\bold r, \omega)$ is the radiation field in the medium, which is produced by the same dipole located in a vacuum far away from the interface (in the wave zone). We assume that dipole moment $\bold d$ is oriented along the normal to the surface through which radiation emerges (i.e., along the $z$ axis in our case). Physically, such an orientation
means that a thin conducting screen at large distances can be treated as a double layer. Taking into account that vector $\bold E^{R}$ is perpendicular to $\bold e$ we can find the modulus of the
radiation field in a vacuum from formula (\ref{Eq21}):
\begin{eqnarray}
\displaystyle && |\bold E^{R(vac)}| = \Big |\frac{\sin \Theta}{\sin \theta}\bold E^{R(m)}\Big | = \Big |\frac{1}{\sqrt{\varepsilon}} \bold E^{R(m)}\Big | = \Big |\frac{1}{\varepsilon}\bold H^{R(m)}\Big |, \label{Eq22}
\end{eqnarray}
Here, we have used the Snell's law connecting ``vacuum'' angle $\theta$ and angle $\Theta$ in the medium, as well as the fact that the following equality holds for the fields of a spherical wave in the medium: $|\bold E^{R(m)}| = |\varepsilon^{-1/2}\bold H^{R(m)}|$. 

The relations between the angles at which radiation is emitted in the medium and in the vacuum, which appear in formula (\ref{Eq16}), can be expressed as follows (compare with (\ref{Eq14})):
\begin{eqnarray}
\displaystyle && \bold e = \frac{1}{\sqrt{\varepsilon }}\{\sin \theta \cos \phi, \ \sin \theta \sin \phi, \sqrt{\varepsilon - \sin^2 \theta}\}. \label{Eq23}
\end{eqnarray}

Then, in accordance with formula (\ref{Eq22}), we must find the magnetic field in the medium in the case when the field of the wave incident on the interface from vacuum can be expressed in terms of field (\ref{Eq16}); i.e., we must solve the inverse problem. In the case of conducting media, one can confine the analysis to conventional Fresnel laws for only one interface (i.e., one can disregard the waves reflected in the substance from the second boundary of the screen, which lies in the $z=-d$ plane). Such an algorithm was used in our previous publication \cite{JL}. In the general case, the field
incident on the interface is not equal to ``forward'' radiation (\ref{Eq16}) because the ``backward'' radiation waves propagating at angles $\Theta > \pi/2$ can also emerge into the vacuum through the front face of the plate after their reflection from the rear face. In addition, refraction of the wave incident from vacuum onto the screen should be considered taking into account possible
multiple re-reflections inside it. Here, an important stipulation concerning the screen thickness $d$ under investigation is due. If the size of the region through which radiation emerges into vacuum is much larger than the screen thickness (thin screen), and the medium has a non-zero absorbance ($\varepsilon^{\prime \prime} \ne 0$), the waves reflected from the upper and lower faces of the plate can be disregarded. Indeed, if the inequality $(b-a) \gg d$ holds, only the waves of Cherenkov radiation with $\Theta \sim \pi/2$ are incident on the upper and lower faces of the screen, which corresponds to $\varepsilon^{\prime} \gg 1$. It was mentioned above that such waves experience total internal reflection and can emerge into the vacuum through the front face of the plate at angles $\theta \sim \pi/2$.
We will return to this question in Section \ref{Sect4}. 

On account of the above arguments, we assume that the wave incident on the interface consists of the field of  ``forward'' radiation (\ref{Eq16}), which will be denoted below by index $(F)$, as well as the field of ``backward'' radiation reflected from the rear face of the screen in the $z= - d$ plane (index $(B)$). The backward radiation field can be determined from formula (\ref{Eq16}) as usual by the inversion of the $z$-projection of the wave vector or by the substitution $\sqrt{\varepsilon - \sin^2 \theta} \rightarrow - \sqrt{\varepsilon - \sin^2 \theta}$. When quantities in relation (\ref{Eq22}) are
squared, expression $|\bold H^{R(m)}|^2$ must be written as the sum of the squared moduli of the field components, one of which is perpendicular to the plane of incidence of the wave on the interface, and the other lies in the plane of incidence. Considering that the magnetic field component lying in the plane of incidence is connected with the electric field component perpendicular to the plane of incidence via $\sqrt{\varepsilon}$, we finally obtain the following relations for the energy emitted into the vacuum (into half-space $z > 0$):
\begin{eqnarray}
\displaystyle && \frac{d^2 W}{d \omega d \Omega} = c r^2 |\bold E^{R(vac)}|^2 =  \frac{d^2 W}{d \omega d \Omega}\Big |_{\perp} +  \frac{d^2 W}{d \omega d \Omega}\Big |_{\parallel} = \cr && \qquad \qquad = \frac{c r^2}{|\varepsilon|^2}\Big (|f_H|^2 |H^{R}_{\perp (F)} + R_H H^{R}_{\perp (B)}|^2 +  |\sqrt{\varepsilon} f_E|^2 |H^{R}_{\parallel (F)} + R_E H^{R}_{\parallel (B)}|^2\Big ). \label{Eq24}
\end{eqnarray}
It should be noted that when the wave (\ref{Eq16}) is incident from vacuum, the amplitude of the spherical wave must be taken without factor $\sqrt{\varepsilon}$ in the exponent. In
expression (\ref{Eq24}), the magnetic field component lying in the incidence plane in the coordinate system used here has the form
\begin{eqnarray}
\displaystyle && H^{R}_{\parallel (F, B)} = \sqrt{(H^{R}_{z (F, B)})^2 + (H^{R}_{x (F, B)} \cos \phi + H^{R}_{y (F, B)} \sin \phi)^2}, \label{Eq25}
\end{eqnarray}
while the magnetic field component perpendicular to the incidence plane has the form
\begin{eqnarray}
\displaystyle && H^{R}_{\perp (F, B)} = H^{R}_{y (F, B)} \cos \phi - H^{R}_{x (F, B)} \sin \phi. \label{Eq26}
\end{eqnarray}
Formula (\ref{Eq24}) contains coefficients $R_{H, E}$ of reflection from the boundary $z = - d$ for waves whose magnetic or electric field is perpendicular to the plane of incidence:
\begin{eqnarray}
\displaystyle R_H = \frac{\sqrt{\varepsilon - \sin^2 \theta} - \varepsilon \cos \theta}{\sqrt{\varepsilon - \sin^2 \theta} + \varepsilon \cos \theta} e^{i 2d\frac{\omega}{c}\sqrt{\varepsilon - \sin^2 \theta}},\ R_E = \frac{\sqrt{\varepsilon - \sin^2 \theta} - \cos \theta}{\sqrt{\varepsilon - \sin^2 \theta} + \cos \theta} e^{i 2d\frac{\omega}{c}\sqrt{\varepsilon - \sin^2 \theta}}, \label{Eq27}
\end{eqnarray}
as well as Fresnel coefficients $f_H$ and $f_E$ connecting the amplitude of the wave incident on the surface of the plate from the vacuum with the amplitude of the wave transmitted through the plate with allowance for multiple re-reflections inside it. These coefficients can easily be calculated using the standard boundary conditions in complete analogy with the case of a single interface (see, for example, \cite{Gr, L}). We give these expressions without derivation:
\begin{eqnarray}
\displaystyle  && f_H = 2 \varepsilon \cos \theta \frac{(\sqrt{\varepsilon - \sin^2 \theta} + \varepsilon \cos \theta) e^{-i d\frac{\omega}{c}\sqrt{\varepsilon - \sin^2 \theta}}}{(\varepsilon \cos \theta + \sqrt{\varepsilon - \sin^2 \theta})^2 e^{- i d\frac{\omega}{c}\sqrt{\varepsilon - \sin^2 \theta}} - (\varepsilon \cos \theta - \sqrt{\varepsilon - \sin^2 \theta})^2 e^{i d\frac{\omega}{c}\sqrt{\varepsilon - \sin^2 \theta}}}, \cr && f_E = 2 \cos \theta \frac{(\sqrt{\varepsilon - \sin^2 \theta} + \cos \theta) e^{-i d\frac{\omega}{c}\sqrt{\varepsilon - \sin^2 \theta}}}{(\cos \theta + \sqrt{\varepsilon - \sin^2 \theta})^2 e^{- i d\frac{\omega}{c}\sqrt{\varepsilon - \sin^2 \theta}} - (\cos \theta - \sqrt{\varepsilon - \sin^2 \theta})^2 e^{i d\frac{\omega}{c}\sqrt{\varepsilon - \sin^2 \theta}}}. \label{Eq28}
\end{eqnarray}
For thick screens, these coefficients are transformed into the conventional expressions for a single interface (when $\varepsilon^{\prime \prime} \ne 0$).

Returning to the orientation of the emitting dipole in formula (\ref{Eq21}), we note that if the $\bold d$ component parallel to the plane through which radiation emerges existed, formula (\ref{Eq22}) would acquire a term $\propto \cos \Theta/\cos \theta \propto \sqrt{\varepsilon -\sin^2 \theta}/\sqrt{\varepsilon}$ containing an additional power $\sqrt{\varepsilon}$ in the numerator. This would ultimately turn the radiation intensity to infinity in the limit of a perfect conductivity of the screen ($\varepsilon^{\prime \prime} \rightarrow \infty$).

It can easily be seen that $H^{R}_{\parallel (F, B)} = 0$ in the azimuthally symmetric problem under investigation, and we obtain, after collecting terms, the following expression for the
energy emitted into vacuum in the forward direction ($\theta < \pi/2$):
\begin{eqnarray}
\displaystyle && \frac{d^2W}{d\omega d\Omega}\Big |_F = \frac{e^2}{\pi^2 c}  \frac{\beta^2 \cos^2{\theta}}{(1 - \beta^2 \cos^2{\theta})^2}\Big (\frac {\omega}{v\gamma}\Big )^2 \times \cr && \Bigg |\frac{(\varepsilon - 1) (1 - \beta^2 (\varepsilon - \sin^2 \theta ))^{-1}}{(\varepsilon \cos \theta + \sqrt{\varepsilon - \sin^2 \theta})^2 e^{- i d\frac{\omega}{c}\sqrt{\varepsilon - \sin^2 \theta}} - (\varepsilon \cos \theta - \sqrt{\varepsilon - \sin^2 \theta})^2 e^{i d\frac{\omega}{c}\sqrt{\varepsilon - \sin^2 \theta}}}\Bigg |^2 \times \cr && \displaystyle \times \Big |(\sqrt{\varepsilon - \sin^2 \theta} - \varepsilon \cos \theta )(1 - \beta \sqrt{\varepsilon - \sin^2{\theta}}) \Big (\sin \theta (1 - \beta^2 + \beta \sqrt{\varepsilon - \sin^2{\theta}})\zeta _1 - \cr && - \gamma^{-1} (\beta \sin^2 \theta - \sqrt{\varepsilon - \sin^2 \theta}) \zeta _2 \Big ) e^{i d \frac{\omega}{c} \sqrt{\varepsilon - \sin^2{\theta}}} + (\sqrt{\varepsilon - \sin^2 \theta} + \varepsilon \cos \theta ) (1 + \beta \sqrt{\varepsilon - \sin^2{\theta}}) \cr && \times \Big (\sin \theta (1 - \beta^2 - \beta \sqrt{\varepsilon - \sin^2{\theta}}) \zeta _1 - \gamma^{-1} (\beta \sin^2 \theta + \sqrt{\varepsilon - \sin^2 \theta}) \zeta _2 \Big ) e^{-i d \frac{\omega}{c} \sqrt{\varepsilon - \sin^2{\theta}}} - \cr && \qquad \qquad - 2 \sqrt{\varepsilon - \sin^2 \theta} \Big (\sin \theta (1 - \beta^2 - \beta^2 (\varepsilon - \sin^2{\theta}) - \beta^3 \varepsilon \cos \theta ) \zeta _1 - \cr && \qquad \qquad \qquad \qquad - \gamma^{-1} \varepsilon (\beta + \cos \theta + \beta^2 \sin^2 \theta \cos \theta ) \zeta _2 \Big ) e^{-i \frac{\omega}{v}d}\Big |^2. \label{Eq29}
\end{eqnarray}
Here, the following notation has been introduced:
\begin{eqnarray}
\displaystyle && \zeta _1 = a J_0\Big (a \frac{\omega}{c} \sin{\theta}\Big ) K_1\Big (a \frac{\omega}{v \gamma}\Big ) - b J_0\Big (b \frac{\omega}{c} \sin{\theta}\Big ) K_1\Big (b \frac{\omega}{v \gamma}\Big ), \cr &&
\zeta_2 = a J_1\Big (a \frac{\omega}{c} \sin{\theta}\Big ) K_0\Big (a \frac{\omega}{v \gamma}\Big ) - b J_1\Big (b \frac{\omega}{c} \sin{\theta}\Big ) K_0\Big (b \frac{\omega}{v \gamma}\Big ). \label{Eq30}
\end{eqnarray}

It should be noted from the very outset that in the limiting case of transition radiation in a plate (slab) of thickness $d$ (for which $a \rightarrow 0, b \rightarrow \infty$) functions $\zeta_1$ and $\zeta_2$ have values $v \gamma/\omega$ and $0$, respectively. In this case, formula (\ref{Eq29}) completely coincides with the well-known Pafomov solution \cite{G-T, B, Term, Gar} (pay attention to the obvious misprint in formula (3.55) from \cite{G-T}, which does not appear in \cite{B, Term, Gar}). For thick screens, the solution obtained in this way is transformed into the well-known Ginzburg-Frank formula for transition radiation generated in a semi-infinite medium.

The pole $|1 - \beta \sqrt{\varepsilon - \sin^2 \theta}| \rightarrow 0$ in the denominator of expression (\ref{Eq29}) corresponds to the condition for Cherenkov radiation in a vacuum. However, this pole
can be eliminated, and in the vicinity of Cherenkov angle, formula (\ref{Eq29}) under the assumption of a transparent medium assumes the form
\begin{eqnarray}
\displaystyle && \frac{d^2W}{d\omega d\Omega}\Big |_{F}\rightarrow \frac{d^2W}{d\omega d\Omega}\Big |_{ChR} = \frac{e^2 \omega^2 c^3}{4 \pi^2 \gamma^2 v^6} (1 + \beta^2 - \varepsilon \beta^2) \cr && \qquad \displaystyle \times \Bigg (\frac{(1 - \varepsilon \sqrt{1 + \beta^2 - \varepsilon \beta^2})\Big (\zeta_1 (2 - \beta^2 ) \sqrt{\beta^2 \varepsilon - 1} + \zeta_2 \gamma^{-1} (2 - \beta^2 \varepsilon ) \Big )}{(1 - \varepsilon \sqrt{1 + \beta^2 - \varepsilon \beta^2})^2 - (1 + \varepsilon \sqrt{1 + \beta^2 - \varepsilon \beta^2})^2 e^{-i 2 d\frac{\omega}{v}}}\Bigg )^{2}. \label{Eq29a}
\end{eqnarray}
Since $\sin \theta_{ChR} = \sqrt{\varepsilon - 1/\beta^2} < 1$, the quantities in the radicand are always positive.

A distinguishing feature of formula (\ref{Eq29}) is that the energy emitted in the backward direction ($\theta > \pi/2$) cannot be obtained by simple substitution $\beta \rightarrow - \beta$. Indeed, for consideration of radiation emerging from the rear face of the screen, we must change the places of indices $(F)$ and $(B)$ in formula (\ref{Eq24}), which leads to the substitution $\beta \sqrt{\varepsilon - \sin^2 \theta}\rightarrow - \beta \sqrt{\varepsilon - \sin^2 \theta}$ in the first two terms in formula (\ref{Eq29}), as well as the substitution $\beta \sin^2 \theta - \sqrt{\varepsilon - \sin^2 \theta}\leftrightarrow \beta \sin^2 \theta + \sqrt{\varepsilon - \sin^2 \theta}$. The latter substitution breaks the symmetry observed in the limiting case of transition radiation. Simple sign reversal of the particle velocity in formula (\ref{Eq29}) is impossible if only in view of the fact that, in this formula, as well as in the initial expression (\ref{Eq12}) for the particle field, the velocity appears in the argument of the Bessel function and it is a positive quantity.

It is convenient to calculate the energy emitted in the backward direction in the same coordinate system (as before, $\theta < \pi/2$ in this case) by changing the direction of the particle velocity, which leads to sign reversal of the $z$-component of the particle field (\ref{Eq12}), as well as to the substitution $z\rightarrow - z$ in the exponent. Analogous calculations lead to the following expression:
\begin{eqnarray}
\displaystyle && \frac{d^2W}{d\omega d\Omega}\Big |_B = \frac{e^2}{\pi^2 c}  \frac{\beta^2 \cos^2{\theta}}{(1 - \beta^2 \cos^2{\theta})^2}\Big (\frac {\omega}{v\gamma}\Big )^2 \times \cr && \Bigg |\frac{(\varepsilon - 1) (1 - \beta^2 (\varepsilon - \sin^2 \theta ))^{-1}}{(\varepsilon \cos \theta + \sqrt{\varepsilon - \sin^2 \theta})^2 e^{- i d\frac{\omega}{c}\sqrt{\varepsilon - \sin^2 \theta}} - (\varepsilon \cos \theta - \sqrt{\varepsilon - \sin^2 \theta})^2 e^{i d\frac{\omega}{c}\sqrt{\varepsilon - \sin^2 \theta}}}\Bigg |^2 \times \cr && \displaystyle \Big |(\sqrt{\varepsilon - \sin^2 \theta} - \varepsilon \cos \theta )(1 + \beta \sqrt{\varepsilon - \sin^2{\theta}}) \Big (\sin \theta (1 - \beta^2 - \beta \sqrt{\varepsilon - \sin^2{\theta}})\zeta _1 - \cr && - \gamma^{-1} (\beta \sin^2 \theta + \sqrt{\varepsilon - \sin^2 \theta}) \zeta _2 \Big ) e^{i d \frac{\omega}{c} \sqrt{\varepsilon - \sin^2{\theta}}} + (\sqrt{\varepsilon - \sin^2 \theta} + \varepsilon \cos \theta ) (1 - \beta \sqrt{\varepsilon - \sin^2{\theta}}) \cr && \times \Big (\sin \theta (1 - \beta^2 + \beta \sqrt{\varepsilon - \sin^2{\theta}}) \zeta _1 - \gamma^{-1} (\beta \sin^2 \theta - \sqrt{\varepsilon - \sin^2 \theta}) \zeta _2 \Big ) e^{-i d \frac{\omega}{c} \sqrt{\varepsilon - \sin^2{\theta}}} - \cr && \qquad \qquad - 2 \sqrt{\varepsilon - \sin^2 \theta} \Big (\sin \theta (1 - \beta^2 - \beta^2 (\varepsilon - \sin^2{\theta}) + \beta^3 \varepsilon \cos \theta ) \zeta _1 - \cr && \qquad \qquad \qquad \qquad - \gamma^{-1} \varepsilon (\beta - \cos \theta - \beta^2 \sin^2 \theta \cos \theta ) \zeta _2 \Big ) e^{i \frac{\omega}{v}d}\Big |^2. \label{Eq31}
\end{eqnarray}
It can be seen that formulas (\ref{Eq29}) and (\ref{Eq31}) are transformed into each other under the substitutions (everywhere except in exponents):
\begin{eqnarray}
\displaystyle && \sqrt{\varepsilon - \sin^2 \theta} \rightarrow - \sqrt{\varepsilon - \sin^2 \theta}, \ \cos{\theta}\rightarrow - \cos {\theta}, \ e^{-i \frac{\omega}{v}d}\rightarrow e^{i \frac{\omega}{v}d}. \label{Eq32}
\end{eqnarray} 
Expression (\ref{Eq31}) for the intensity of backward radiation is distinguished by the presence of the Cherenkov pole $|1 - \beta \sqrt{\varepsilon - \sin^2 \theta}| \rightarrow 0$, which does not appear for backward transition radiation generated in a semi-infinite medium. This is due to the fact that radiation emitted in the direction of the particle velocity can be reflected from the front
face of the screen and emerge into the vacuum though the rear face.

In the special case of ideal conductivity ($\varepsilon^{\prime \prime}\rightarrow \infty$), the dependence on screen thickness $d$ disappears (skin effect), and we have the following expression for diffraction radiation from a circular aperture in an infinitely thin screen of a finite radius:
\begin{eqnarray}
\displaystyle && \frac{d^2W}{d\omega d\Omega}\Big |_F = \frac{d^2W}{d\omega d\Omega}\Big |_B = \frac{e^2}{\pi^2 c}  \frac{\beta^2 \sin^2{\theta}}{(1 - \beta^2 \cos^2{\theta})^2}\Big (\frac {\omega}{v\gamma}\Big )^2 \Big (\zeta_1 + \frac{\zeta_2}{\beta \gamma \sin \theta}\Big )^2,  \label{Eq33}
\end{eqnarray}
Note that the energy emitted in both directions is the same in this case. This solution coincides with that obtained earlier in \cite{J} using another method and (in the ultrarelativistic
case) with the results obtained in \cite{Mono, Dn, Bol, X}, where the problem was solved in the ultrarelativistic limit from the very outset. Since the dependence on the screen thickness has disappeared, formula (\ref{Eq33}) is also valid for a non-relativistic charge if radiation losses can still be regarded as small as compared to the total energy of the particle.

It should be emphasized that consideration of radiation emerging from the screen into vacuum through the $z = 0$ plane corresponds to a small screen thickness and, accordingly, to the emergence of the skin effect for $\varepsilon^{\prime \prime} \gg 1$ exactly in the vicinity of this plane. For this reason, the transition to radiation in a perfectly conducting waveguide, as well as the transition to Cherenkov radiation in an infinitely long channel, cannot be made using the formulas provided.

\section{\large{Polarization Radiation from a Rectangular Screen}}
\label{Sect2}

In the case of azimuthal symmetry of the problem, the terms with coefficient $f_E$ make zero contribution to general expression (\ref{Eq24}) for the emitted energy because the magnetic field is polarized perpendicularly to the plane of incidence. If polarization radiation is generated by a particle moving near a rectangular screen (see Fig. 2), this circumstance does not take place
because the azimuthal symmetry is broken. In the coordinate system used here, the magnetic field components of radiation have the form
\begin{eqnarray}
\displaystyle && H^{R}_{\perp (F, B)} = H^{R}_{x (F, B)} \cos \phi - H^{R}_{y (F, B)} \sin \phi, \cr && H^{R}_{\parallel (F, B)} = \sqrt{(H^{R}_{z (F, B)})^2 + (H^{R}_{x (F, B)} \sin \phi + H^{R}_{y (F, B)} \cos \phi)^2}. \label{Eq34}
\end{eqnarray}

To find the radiation field, we take advantage of the fact that the screen width along the $x$ axis is infinitely large. Then formula (\ref{Eq11}) gives
\begin{eqnarray}
\displaystyle && \bold H^{R} (\bold r, \omega ) = \frac{2 \pi i \omega}{c^2} \sqrt{\varepsilon} \frac{e^{i r \sqrt{\varepsilon} \omega/c}}{r} \bold e \times \int \limits_{-d}^0 dz^{\prime} \int \limits_0^a dy^{\prime} \sigma (\omega) \bold E^{0} (k_x, y^{\prime}, z^{\prime}, \omega) e^{-i k_y y^{\prime} - i k_z z^{\prime}}. \label{Eq35}
\end{eqnarray}
The corresponding Fourier component of the field produced by the charge has the form
\begin{eqnarray}
\displaystyle \bold E^{0} (k_x, y^{\prime}, z^{\prime}, \omega) = \frac{- i e}{2 \pi v} \frac{e^{i \frac{\omega}{v}z^{\prime}}}{\sqrt{1 + \varepsilon (\beta \gamma e_x)^2}}
\{ \sqrt{\varepsilon}\beta\gamma e_x, i \sqrt{1 + \varepsilon (\beta \gamma e_x)^2}, \gamma^{-1}\} \ e^{-(y^{\prime} + h) \frac{\omega}{v \gamma}
\sqrt{1 + \varepsilon (\beta \gamma e_x)^2}} \label{Eq36}
\end{eqnarray}
Here, $h$ is the distance between the particle trajectory and the screen, and unit vector $\bold e$ of radiation in the ``vacuum variables'' has the form
\begin{eqnarray}
\displaystyle && \bold {e} = \frac{1}{\sqrt{\varepsilon }} \{\sin{\theta} \sin {\phi}, \sin{\theta} \cos {\phi}, \sqrt{\varepsilon - \sin^2 \theta}\}. \label{Eq37}
\end{eqnarray}
Substituting expression (\ref{Eq36}) into (\ref{Eq35}) and (\ref{Eq34}), we obtain the field components of radiation:
\begin{eqnarray}
\displaystyle && H^{R}_{\perp (F, B)} = \frac{e}{\pi c} \frac{\beta \gamma}{4}(\varepsilon -1) \frac{e^{i r \sqrt{\varepsilon} \omega/c}}{r} \frac{1 - e^{i d \frac{\omega}{c}
(-\beta^{-1} \pm \sqrt{\varepsilon - \sin^2{\theta}})}}{1 \mp \beta \sqrt{\varepsilon - \sin^2{\theta}}}e^{-h \frac{\omega}{v \gamma} \sqrt{1 + (\beta \gamma \sin \theta \sin \phi)^2}}\cr && \qquad \times \displaystyle \frac{e^{-a \frac{\omega}{v \gamma} (i \beta \gamma \sin \theta \cos \phi + \sqrt{1 + (\beta \gamma \sin \theta \sin \phi )^2})} - 1}{\sqrt{1 + (\beta \gamma \sin \theta \sin \phi )^2} (i \beta \gamma \sin \theta \cos \phi + \sqrt{1 + (\beta \gamma \sin \theta \sin \phi)^2})} \Big (\gamma^{-1} \sin \theta \mp \cr \displaystyle && \qquad \qquad \qquad \qquad \sqrt{\varepsilon - \sin^2 \theta} (i \cos \phi \sqrt{1 + (\beta \gamma \sin \theta \sin \phi )^2} + \beta \gamma \sin \theta \sin^2 \phi ) \Big ), \cr 
\displaystyle && H^{R}_{\parallel (F, B)} = \frac{e}{\pi c} \frac{\beta \gamma}{4}(\varepsilon -1) \frac{e^{i r \sqrt{\varepsilon} \omega/c}}{r} \frac{1 - e^{i d \frac{\omega}{c}
(-\beta^{-1} \pm \sqrt{\varepsilon - \sin^2{\theta}})}}{1 \mp \beta \sqrt{\varepsilon - \sin^2{\theta}}}e^{-h \frac{\omega}{v \gamma} \sqrt{1 + (\beta \gamma \sin \theta \sin \phi)^2}}\cr \displaystyle && \qquad \times \frac{e^{-a \frac{\omega}{v \gamma} (i \beta \gamma \sin \theta \cos \phi + \sqrt{1 + (\beta \gamma \sin \theta \sin \phi )^2})} - 1}{\sqrt{1 + (\beta \gamma \sin \theta \sin \phi )^2} (i \beta \gamma \sin \theta \cos \phi + \sqrt{1 + (\beta \gamma \sin \theta \sin \phi)^2})} \cr && \qquad \qquad \qquad \times \sqrt{\varepsilon} \sin \phi \Big (\beta \gamma \sin \theta \cos \phi - i \sqrt{1 + (\beta \gamma \sin \theta \sin \phi )^2} \Big ). \label{Eq38}
\end{eqnarray}
The upper sign in these expressions corresponds to forward radiation and the lower sign, to backward radiation. In raising these expressions to the second power, we can use the relation
\begin{eqnarray}
\displaystyle & \Big |\exp\Big \{-a \frac{\omega}{v \gamma} (i \beta \gamma \sin \theta \cos \phi + \sqrt{1 + (\beta \gamma \sin \theta \sin \phi )^2})\Big \} - 1\Big |^2 = \cr \displaystyle & = 4 \Big (\sinh^2\Big (\frac{a}{2}\frac{\omega}{v \gamma} \sqrt{1 + (\beta \gamma \sin \theta \sin \phi )^2}\Big ) + \sin^2 \Big (\frac{a}{2}\frac{\omega}{c} \sin \theta \cos \phi\Big ) \Big ) e^{-a \frac{\omega}{v \gamma} \sqrt{1 + (\beta \gamma \sin \theta \sin \phi )^2}}. \label{Eq39}
\end{eqnarray}
Substituting expressions (\ref{Eq38}) into (\ref{Eq24}) and collecting terms, we obtain for forward radiation
\begin{eqnarray}
\displaystyle && \frac{d^2W}{d\omega d\Omega}\Big |_{\perp (F)} = \frac{e^2}{\pi^2 c}  \frac{\beta^2 \cos^2{\theta}}{(1 - \beta^2 \cos^2{\theta}) (1 + (\beta \gamma \sin \theta \sin \phi )^2)}\Big (\sinh^2 \Big (\frac{\omega}{v \gamma} \frac{a}{2} \sqrt{1 + (\beta \gamma \sin \theta \sin \phi )^2}\Big ) \cr && + \sin^2 \Big (\frac{\omega}{c} \frac{a}{2} \sin \theta \cos \phi \Big ) \Big ) \Big |\frac{\varepsilon - 1}{1 - \beta^2 (\varepsilon - \sin^2 \theta )} \Big |^2 e^{-(h + \frac{a}{2}) \frac{2 \omega}{v \gamma} \sqrt{1 + (\beta \gamma \sin \theta \sin \phi )^2}}\cr &&  \times \Big |(\varepsilon \cos \theta + \sqrt{\varepsilon - \sin^2 \theta})^2 e^{- i d\frac{\omega}{c}\sqrt{\varepsilon - \sin^2 \theta}} - (\varepsilon \cos \theta - \sqrt{\varepsilon - \sin^2 \theta})^2 e^{i d\frac{\omega}{c}\sqrt{\varepsilon - \sin^2 \theta}}\Big |^{-2} \cr &&  \times \Big | (\sqrt{\varepsilon - \sin^2 \theta} - \varepsilon \cos \theta )(1 - \beta \sqrt{\varepsilon - \sin^2{\theta}}) \Big (\gamma^{-1}\sin \theta + \sqrt{\varepsilon - \sin^2{\theta}}(\beta \gamma \sin \theta \sin^2 \phi + \cr && + i\cos \phi \sqrt{1 + (\beta \gamma \sin \theta \sin \phi )^2}) \Big ) e^{i d \frac{\omega}{c} \sqrt{\varepsilon - \sin^2{\theta}}} + (\sqrt{\varepsilon - \sin^2 \theta} + \varepsilon \cos \theta )(1 + \beta \sqrt{\varepsilon - \sin^2{\theta}}) \cr && \Big (\gamma^{-1}\sin \theta - \sqrt{\varepsilon - \sin^2{\theta}}(\beta \gamma \sin \theta \sin^2 \phi + i\cos \phi \sqrt{1 + (\beta \gamma \sin \theta \sin \phi )^2}) \Big ) e^{-i d \frac{\omega}{c} \sqrt{\varepsilon - \sin^2{\theta}}} - \cr && \qquad \qquad - 2 \sqrt{\varepsilon - \sin^2 \theta} \Big (\gamma^{-1}\sin \theta (1 + \beta \varepsilon \cos \theta ) - (\varepsilon \cos \theta + \beta (\varepsilon - \sin^2 \theta )) (\beta \gamma \sin \theta \sin^2 \phi + \cr && \qquad \qquad \qquad \qquad + i \cos \phi \sqrt{1 + (\beta \gamma \sin \theta \sin \phi )^2}) \Big )
e^{-i \frac{\omega}{v}d}\Big |^2, \cr && \displaystyle \frac{d^2W}{d\omega d\Omega}\Big |_{\parallel (F)} = \frac{e^2}{\pi^2 c}  \frac{(\beta \gamma \cos{\theta} \sin \phi )^2}{1 + (\beta \gamma \sin \theta \sin \phi )^2}\Big (\sinh^2 \Big (\frac{\omega}{v \gamma} \frac{a}{2} \sqrt{1 + (\beta \gamma \sin \theta \sin \phi )^2}\Big ) \cr && + \sin^2 \Big (\frac{\omega}{c} \frac{a}{2} \sin \theta \cos \phi \Big ) \Big ) \Big |\frac{\varepsilon - 1}{1 - \beta^2 (\varepsilon - \sin^2 \theta )} \Big |^2 e^{-(h + \frac{a}{2}) \frac{2 \omega}{v \gamma} \sqrt{1 + (\beta \gamma \sin \theta \sin \phi )^2}}\cr && \times \Big |(\cos \theta + \sqrt{\varepsilon - \sin^2 \theta})^2 e^{- i d\frac{\omega}{c}\sqrt{\varepsilon - \sin^2 \theta}} - (\cos \theta - \sqrt{\varepsilon - \sin^2 \theta})^2 e^{i d\frac{\omega}{c}\sqrt{\varepsilon - \sin^2 \theta}}\Big |^{-2} \cr && \ \times \Big | (\sqrt{\varepsilon - \sin^2 \theta} - \cos \theta )(1 - \beta \sqrt{\varepsilon - \sin^2{\theta}}) e^{i d \frac{\omega}{c} \sqrt{\varepsilon - \sin^2{\theta}}} + \cr && \qquad + (\sqrt{\varepsilon - \sin^2 \theta} + \cos \theta )(1 + \beta \sqrt{\varepsilon - \sin^2{\theta}}) e^{-i d \frac{\omega}{c} \sqrt{\varepsilon - \sin^2{\theta}}} - \cr && \qquad \qquad \qquad \qquad - 2 \sqrt{\varepsilon - \sin^2 \theta} (1 + \beta \cos \theta ) e^{-i \frac{\omega}{v}d}\Big |^2. \label{Eq40}
\end{eqnarray}
The total intensity is given by the sum of the above expressions. It should be noted that the decomposition of the spectral-angular density into components used here is determined by the polarization characteristics of the magnetic field of radiation (and not by the electric field, as is often used; see, for example, \cite{B}). 

A distinguishing feature of the above expressions is allowance for possible multiple re-reflections of radiation inside the screen. In the case of a conducting screen, the contribution from radiation reflected from the rear face of the plate is small, and the formula derived for such a screen coincides with that obtained in \cite{JL}. It should also be noted that the resultant formula
contains the Cherenkov pole $|1 - \beta \sqrt{\varepsilon - \sin^2 \theta}| \rightarrow 0$, which can be eliminated for a finite width $d$ of the screen.

It should be emphasized that, to obtain a formula for backward radiation, it is insufficient to change places of the corresponding fields in expression (\ref{Eq24}) because in this case we consider the emergence of radiation into vacuum through the $z= - d$ plane and not through the plane $z = 0$. To obtain the correct result, we must also change the direction of the particle velocity, which corresponds to sign reversal for the $z$-component and the substitution $e^{i \frac{\omega}{v}z^{\prime}}\rightarrow e^{-i \frac{\omega}{v}z^{\prime}}$ in formula (\ref{Eq36}). 
In this case, the backward radiation field propagating in the positive direction of the $z$ axis differs from formula (\ref{Eq38}) in the substitution $-\beta^{-1} - \sqrt{\varepsilon - \sin^2{\theta}} \rightarrow \beta^{-1} + \sqrt{\varepsilon - \sin^2{\theta}}$ in the exponent. Taking into account these arguments, we obtain the following expression for the energy emitted in the backward direction (as before, $\theta < \pi/2$):
\begin{eqnarray}
\displaystyle && \frac{d^2W}{d\omega d\Omega}\Big |_{\perp (B)} = \frac{e^2}{\pi^2 c}  \frac{\beta^2 \cos^2{\theta}}{(1 - \beta^2 \cos^2{\theta}) (1 + (\beta \gamma \sin \theta \sin \phi )^2)}\Big (\sinh^2 \Big (\frac{\omega}{v \gamma} \frac{a}{2} \sqrt{1 + (\beta \gamma \sin \theta \sin \phi )^2}\Big ) \cr && + \sin^2 \Big (\frac{\omega}{c} \frac{a}{2} \sin \theta \cos \phi \Big ) \Big ) \Big |\frac{\varepsilon - 1}{1 - \beta^2 (\varepsilon - \sin^2 \theta )} \Big |^2 e^{-(h + \frac{a}{2}) \frac{2 \omega}{v \gamma} \sqrt{1 + (\beta \gamma \sin \theta \sin \phi )^2}}\cr &&  \times \Big |(\varepsilon \cos \theta + \sqrt{\varepsilon - \sin^2 \theta})^2 e^{- i d\frac{\omega}{c}\sqrt{\varepsilon - \sin^2 \theta}} - (\varepsilon \cos \theta - \sqrt{\varepsilon - \sin^2 \theta})^2 e^{i d\frac{\omega}{c}\sqrt{\varepsilon - \sin^2 \theta}}\Big |^{-2} \cr &&  \times \Big | (\sqrt{\varepsilon - \sin^2 \theta} - \varepsilon \cos \theta )(1 + \beta \sqrt{\varepsilon - \sin^2{\theta}}) \Big (\gamma^{-1}\sin \theta - \sqrt{\varepsilon - \sin^2{\theta}}(\beta \gamma \sin \theta \sin^2 \phi + \cr && + i\cos \phi \sqrt{1 + (\beta \gamma \sin \theta \sin \phi )^2}) \Big ) e^{i d \frac{\omega}{c} \sqrt{\varepsilon - \sin^2{\theta}}} + (\sqrt{\varepsilon - \sin^2 \theta} + \varepsilon \cos \theta )(1 - \beta \sqrt{\varepsilon - \sin^2{\theta}}) \cr && \Big (\gamma^{-1}\sin \theta + \sqrt{\varepsilon - \sin^2{\theta}}(\beta \gamma \sin \theta \sin^2 \phi + i\cos \phi \sqrt{1 + (\beta \gamma \sin \theta \sin \phi )^2}) \Big ) e^{-i d \frac{\omega}{c} \sqrt{\varepsilon - \sin^2{\theta}}} - \cr && \qquad \qquad - 2 \sqrt{\varepsilon - \sin^2 \theta} \Big (\gamma^{-1}\sin \theta (1 - \beta \varepsilon \cos \theta ) - (-\varepsilon \cos \theta + \beta (\varepsilon - \sin^2 \theta )) (\beta \gamma \sin \theta \sin^2 \phi + \cr && \qquad \qquad \qquad \qquad + i \cos \phi \sqrt{1 + (\beta \gamma \sin \theta \sin \phi )^2}) \Big )
e^{i \frac{\omega}{v}d}\Big |^2, \cr && \displaystyle \frac{d^2W}{d\omega d\Omega}\Big |_{\parallel (B)} = \frac{e^2}{\pi^2 c}  \frac{(\beta \gamma \cos{\theta} \sin \phi )^2}{1 + (\beta \gamma \sin \theta \sin \phi )^2}\Big (\sinh^2 \Big (\frac{\omega}{v \gamma} \frac{a}{2} \sqrt{1 + (\beta \gamma \sin \theta \sin \phi )^2}\Big ) \cr && + \sin^2 \Big (\frac{\omega}{c} \frac{a}{2} \sin \theta \cos \phi \Big ) \Big ) \Big |\frac{\varepsilon - 1}{1 - \beta^2 (\varepsilon - \sin^2 \theta )} \Big |^2 e^{-(h + \frac{a}{2}) \frac{2 \omega}{v \gamma} \sqrt{1 + (\beta \gamma \sin \theta \sin \phi )^2}}\cr && \times \Big |(\cos \theta + \sqrt{\varepsilon - \sin^2 \theta})^2 e^{- i d\frac{\omega}{c}\sqrt{\varepsilon - \sin^2 \theta}} - (\cos \theta - \sqrt{\varepsilon - \sin^2 \theta})^2 e^{i d\frac{\omega}{c}\sqrt{\varepsilon - \sin^2 \theta}}\Big |^{-2} \cr && \ \times \Big | (\sqrt{\varepsilon - \sin^2 \theta} - \cos \theta )(1 + \beta \sqrt{\varepsilon - \sin^2{\theta}}) e^{i d \frac{\omega}{c} \sqrt{\varepsilon - \sin^2{\theta}}} + \cr && \qquad + (\sqrt{\varepsilon - \sin^2 \theta} + \cos \theta )(1 - \beta \sqrt{\varepsilon - \sin^2{\theta}}) e^{-i d \frac{\omega}{c} \sqrt{\varepsilon - \sin^2{\theta}}} - \cr && \qquad \qquad \qquad \qquad - 2 \sqrt{\varepsilon - \sin^2 \theta} (1 - \beta \cos \theta ) e^{i \frac{\omega}{v}d}\Big |^2. \label{Eq41}
\end{eqnarray}
It can easily be seen that formulas (\ref{Eq40}) and (\ref{Eq41}) are transformed into each other upon substitution (\ref{Eq32}) as well as upon the substitution $\beta \rightarrow -\beta, \ \cos \phi \rightarrow - \cos \phi$, which corresponds to a shift in the azimuthal angle by $\pi$. This expression also contains the Cherenkov pole associated with the contribution of waves reflected from the front face of the screen and emerging into the vacuum through the rear face.

In the special case of a perfectly conducting semi-plane ($\varepsilon^{\prime \prime} \rightarrow \infty, \ a \rightarrow \infty$), the energy emitted into both half-spaces is also the same and given by
\begin{eqnarray}
\displaystyle \frac{d^2 W}{d\omega d \Omega} = \frac{e^2}{4 \pi^2 c} \frac{1 - \sin^2{\theta}\sin^2{\phi} + (\beta \gamma \sin \theta \sin \phi )^2 (1 + \cos^2{\theta})}{ (1 - \beta^2 \cos^2 \theta ) (1 + (\beta \gamma \sin\theta \sin \phi )^2)} e^{- h \frac{2 \omega}{v \gamma} \sqrt{1 + (\beta \gamma \sin \theta \sin \phi)^2}}, \label{Eq42}
\end{eqnarray}
which coincides with the result obtained in \cite{J, PLA} using another approach.

Comparison of the above formulas with the results obtained in the previous section shows that the lack of azimuthal symmetry leads to the emergence of factor $1 + (\beta \gamma \sin\theta \sin \phi )^2$ in the denominator of the formulas for the emitted energy, which gives in the relativistic case a radiation peak in the vicinity of the $yz$ plane (perpendicular to the plane of the screen) for angles $\theta \gg \gamma^{-1}$ (i.e., in fact, for Cherenkov radiation; see also \cite{JL, Mono}). In view of the absence of factor $\sin^2 \theta$ in the numerators of formulas (\ref{Eq40}) and (\ref{Eq41}), the peak in the angular distribution of diffraction radiation corresponds to angle $\theta = 0$, as expected. The absence of the ``Lorentz term'' $1 - \beta^2 \cos^2 \theta$ in the denominator of the
expressions for the parallel component of intensity $d^2W/d\omega d\Omega |_{\parallel}$ is also worth noting.

\section{\large{Polarization Radiation form a Thin Grating of a Finite Permittivity}}
\label{Sect3}

The method of induced currents being developed in this paper can also be employed for calculating radiation emitted by atoms and molecules of a grating consisting of $N$ strips with a rectangular cross section and a finite permittivity $\varepsilon (\omega) = \varepsilon^{\prime} + i \varepsilon^{\prime \prime}$ under the action of the field of a moving particle (so-called Smith-Purcell radiation; see Fig. 3). To simplify calculations, we can require, analogously to the previous section, that thickness $b$ of the strips be much smaller than the strip length $d - a$:
\begin{eqnarray}
\displaystyle b \ll d - a, \label{Eq42b}
\end{eqnarray}
where $d$ is the grating period and $a$ is the width of the vacuum gap between the strips. In this case, we can assume that radiation emerges into the vacuum only through the upper or lower faces of the strips. Since large emission angles $\vartheta$ are mainly of practical interest for Smith-Purcell radiation, the thin-grating approximation employed here can be used for describing actual experiments in the relevant field.

However, in the geometry considered here, in addition to the thin-grating approximation, we must require that the width of the vacuum gaps between the strips be much smaller than the radiation wavelength. For a large number of periods $N \gg 1$ the radiation wavelength is rigidly connected with the grating period (see below); therefore, for all angles $\vartheta $ far from zero and $\pi$, this condition has the form
\begin{eqnarray}
\displaystyle a \ll \lambda \sim d. \label{Eq43}
\end{eqnarray}
This condition stems from the fact that while multiple re-reflections of radiation in each strip can be generally taken into account, re-reflections of radiation in the vacuum gaps cannot be accounted for in the given approach. If, however, the strips are quite thin ($b \ll d$), allowance for multiple re-reflections of radiation between the strips would have led to corrections in the range of angles $\vartheta \sim b/d \ll 1$ even if condition (\ref{Eq43}) were violated. For this reason, we will assume in further analysis that the following inequalities are satisfied:
\begin{eqnarray}
\displaystyle \pi - b/d > \vartheta > b/d. \label{Eq44}
\end{eqnarray}

It will be clear from further analysis that condition (\ref{Eq43}) essentially leads to the representation of the radiation field from the grating in the form of the vector sum of the radiation fields from each strip, which is obviously true only when re-reflections of radiation between the strips are ignored and for the angles satisfying inequalities (\ref{Eq44}). In the approach developed here
(which takes into account the actual dielectric properties of the grating), summation is performed for radiation fields \textit{inside} the medium and not in the vacuum, which exactly corresponds to small scattering of radiation by the "particles" with a size $a \ll \lambda$. Finally, formulas (\ref{Eq24}) for the radiation intensity, as well as formulas (\ref{Eq28}) for the Fresnel coefficients derived for an infinitely large plate, can be used only under conditions (\ref{Eq42b}) and (\ref{Eq43}). This circumstance imposes substantial constraints on the domain of applicability for a number of well-known models of radiation, in which calculations were performed from the very outset for a perfectly conducting grating (see Section 5 below).

Let us first consider radiation above the grating (i.e., in the half-space in which the charge moves). It is convenient to solve the problem in variables $\theta, \phi$ used in the previous section (components of the unit vector of radiation are given in expression (\ref{Eq37})). A transition to variables $\vartheta, \varphi$ which are conventionally used in the Smith-Purcell radiation geometry can be
made in the final formulas. The expression for the required Fourier component of the field of the charge in the ``vacuum variables`` has the form
\begin{eqnarray}
\displaystyle && \bold E^{0} (k_x, y^{\prime}, z^{\prime}, \omega) = \frac{- i e}{2 \pi v} \frac{e^{i \frac{\omega}{v}y^{\prime}}}{\sqrt{1 + (\beta \gamma \sin \theta \sin \phi)^2}}
\{\beta\gamma \sin \theta \sin \phi, \gamma^{-1},  \cr \displaystyle && \qquad \qquad \qquad \qquad \qquad - i \sqrt{1 + (\beta \gamma \sin \theta \sin \phi)^2}\} e^{(z^{\prime} - h) \frac{\omega}{v \gamma} \sqrt{1 + (\beta \gamma \sin \theta \sin \phi)^2}}. \label{Eq45}
\end{eqnarray}
Here, $h$ is the distance between the particle trajectory and the grating. The radiation field inside the grating can be determined analogously to formula (\ref{Eq35}) (the grating width along the $x$ axis is assumed to be infinitely large), the only difference being that the integration is carried out over the volume of all $N$ strips. The field components required for calculating the energy emitted into the vacuum are then determined using formulas (\ref{Eq34}). Taking into account the well-known equality (calculation is performed for an odd number of strips, but the final formulas for the radiation intensity will also be applicable for even values of $N$)
\begin{eqnarray}
\displaystyle && \sum \limits_{n = -\frac{N-1}{2}}^{\frac{N-1}{2}} \exp\Big \{i n d \frac{\omega}{c}\Big (\beta^{-1} - \sin \theta \cos \phi \Big )\Big \} = \frac{\sin \Big (N \frac{d}{2} \frac{\omega}{c}\Big (\frac{1}{\beta} - \sin \theta \cos \phi \Big )\Big )}{\sin \Big (\frac{d}{2} \frac{\omega}{c}\Big (\frac{1}{\beta} - \sin \theta \cos \phi \Big )\Big )} \label{Eq46}
\end{eqnarray}
we obtain the magnetic field components of radiation in the form
\begin{eqnarray}
\displaystyle && H^{R}_{\perp (F, B)} =  \frac{i e}{2 \pi c} \beta \gamma (\varepsilon -1) \frac{e^{i r \sqrt{\varepsilon} \omega/c}}{r} \Big (i \sin \theta \sqrt{1 + (\beta \gamma \sin \theta \sin \phi)^2} \pm \sqrt{\varepsilon - \sin^2 \theta}\times \cr &&  \displaystyle (\gamma^{-1} \cos \phi + \beta \gamma \sin \theta \sin^2 \phi )\Big ) \frac{1 - e^{-b \frac{\omega}{v \gamma} \Big (\sqrt{1 + (\beta \gamma \sin \theta \sin \phi )^2} \mp i \beta \gamma \sqrt{\varepsilon - \sin^2 \theta}\Big )}}{\sqrt{1 + (\beta \gamma \sin \theta \sin \phi )^2} \Big (\sqrt{1 + (\beta \gamma \sin \theta \sin \phi )^2} \mp i \beta \gamma \sqrt{\varepsilon - \sin^2 \theta}\Big )} \cr &&  \displaystyle \qquad \times \frac{\sin \Big (\frac{d-a}{2} \frac{\omega}{c} (\beta^{-1} - \sin \theta \cos \phi )\Big )}{1 - \beta \sin \theta \cos \phi } \frac{\sin \Big (N \frac{d}{2} \frac{\omega}{c}\Big (\frac{1}{\beta} - \sin \theta \cos \phi \Big )\Big )}{\sin \Big (\frac{d}{2} \frac{\omega}{c}\Big (\frac{1}{\beta} - \sin \theta \cos \phi \Big )\Big )} e^{-h \frac{\omega}{v \gamma} \sqrt{1 + (\beta \gamma \sin \theta \sin \phi)^2}}, 
\cr 
\displaystyle && H^{R}_{\parallel (F, B)} =  \frac{i e}{2 \pi c} \beta \gamma (\varepsilon -1) \frac{e^{i r \sqrt{\varepsilon} \omega/c}}{r} \sqrt{\varepsilon} \sin \phi (\gamma^{-1} - \beta \gamma \sin \theta \cos \phi ) \times \cr &&  \displaystyle \qquad \qquad \frac{1 - e^{-b \frac{\omega}{v \gamma} \Big (\sqrt{1 + (\beta \gamma \sin \theta \sin \phi )^2} \mp i \beta \gamma \sqrt{\varepsilon - \sin^2 \theta}\Big )}}{\sqrt{1 + (\beta \gamma \sin \theta \sin \phi )^2} \Big (\sqrt{1 + (\beta \gamma \sin \theta \sin \phi )^2} \mp i \beta \gamma \sqrt{\varepsilon - \sin^2 \theta}\Big )} \cr &&  \displaystyle \qquad \times \frac{\sin \Big (\frac{d-a}{2} \frac{\omega}{c} (\beta^{-1} - \sin \theta \cos \phi )\Big )}{1 - \beta \sin \theta \cos \phi } \frac{\sin \Big (N \frac{d}{2} \frac{\omega}{c}\Big (\frac{1}{\beta} - \sin \theta \cos \phi \Big )\Big )}{\sin \Big (\frac{d}{2} \frac{\omega}{c}\Big (\frac{1}{\beta} - \sin \theta \cos \phi \Big )\Big )} e^{-h \frac{\omega}{v \gamma} \sqrt{1 + (\beta \gamma \sin \theta \sin \phi)^2}}. \label{Eq47}
\end{eqnarray}
Here, the upper sign corresponds to a wave propagating in the positive direction of the $z$ axis (subscript F), while the lower sign corresponds to the negative direction (subscript B). 
The same notations as in the previous sections are used for convenience. It can be seen from these formulas that the number of grating periods appears only in the interference term of the
form $\sin (N x)/\sin (x)$, which corresponds to vector summation of the radiation field from each strip. Substitution of the radiation field components into formula (\ref{Eq24}) for the intensity leads to the following expressions for radiation into the upper half-space:
\begin{eqnarray}
\displaystyle && \frac{d^2W}{d\omega d\Omega}\Big |_{\perp (up)} = \frac{e^2}{\pi^2 c}  \frac{(\beta \gamma )^{-2} \cos^2{\theta}}{1 + (\beta \gamma \sin \theta \sin \phi )^2} \Big |\frac{\varepsilon - 1}{\varepsilon - (\sin \theta \cos \phi )^2} \Big |^2 \frac{\sin^2 \Big (\frac{d-a}{2} \frac{\omega}{c} (\beta^{-1} - \sin \theta \cos \phi )\Big )}{(1 - \beta \sin \theta \cos \phi )^2}\times \cr && \frac{\sin^2 \Big (N \frac{d}{2} \frac{\omega}{c}\Big (\frac{1}{\beta} - \sin \theta \cos \phi \Big )\Big )}{\sin^2 \Big (\frac{d}{2} \frac{\omega}{c}\Big (\frac{1}{\beta} - \sin \theta \cos \phi \Big )\Big )} e^{-h \frac{2 \omega}{v \gamma} \sqrt{1 + (\beta \gamma \sin \theta \sin \phi )^2}} \Big |(\varepsilon \cos \theta + \sqrt{\varepsilon - \sin^2 \theta})^2 e^{- i b\frac{\omega}{c}\sqrt{\varepsilon - \sin^2 \theta}} - \cr && - (\varepsilon \cos \theta - \sqrt{\varepsilon - \sin^2 \theta})^2 e^{i b\frac{\omega}{c}\sqrt{\varepsilon - \sin^2 \theta}}\Big |^{-2}\times \Big | (\sqrt{\varepsilon - \sin^2 \theta} - \varepsilon \cos \theta )\Big (\sqrt{1 + (\beta \gamma \sin \theta \sin \phi )^2} - \cr && - i \beta \gamma \sqrt{\varepsilon - \sin^2{\theta}}\Big ) \Big (i \sin \theta \sqrt{1 + (\beta \gamma \sin \theta \sin \phi )^2} - \sqrt{\varepsilon - \sin^2{\theta}}(\gamma^{-1} \cos \phi + \beta \gamma \sin \theta \sin^2 \phi ) \Big ) \cr && \times e^{i b \frac{\omega}{c} \sqrt{\varepsilon - \sin^2{\theta}}} + (\sqrt{\varepsilon - \sin^2 \theta} + \varepsilon \cos \theta )\Big (\sqrt{1 + (\beta \gamma \sin \theta \sin \phi )^2} + i \beta \gamma \sqrt{\varepsilon - \sin^2{\theta}}\Big ) \times \cr && \Big (i \sin \theta \sqrt{1 + (\beta \gamma \sin \theta \sin \phi )^2} + \sqrt{\varepsilon - \sin^2{\theta}}(\gamma^{-1} \cos \phi + \beta \gamma \sin \theta \sin^2 \phi ) \Big ) e^{-i b \frac{\omega}{c} \sqrt{\varepsilon - \sin^2{\theta}}} - \cr && \quad - 2 \sqrt{\varepsilon - \sin^2 \theta} \Big (i \beta \cos \phi (\varepsilon - \sin^2 \theta) + i \sin \theta (1 + \varepsilon(\beta \gamma \sin \phi )^2) + \varepsilon \cos \theta \times \cr && \qquad \qquad \sqrt{1 + (\beta \gamma \sin \theta \sin \phi )^2} (\gamma^{-1} \cos \phi - \beta \gamma \sin \theta \cos^2 \phi )\Big )
e^{- b \frac{\omega}{v \gamma} \sqrt{1 + (\beta \gamma \sin \theta \sin \phi )^2}}\Big |^2, \cr &&
\displaystyle \frac{d^2W}{d\omega d\Omega}\Big |_{\parallel (up)} = \frac{e^2}{\pi^2 c}  \frac{(\beta \gamma )^{-2} \cos^2{\theta}}{1 + (\beta \gamma \sin \theta \sin \phi )^2} \sin^2 \phi (\gamma^{-1} - \beta \gamma \sin \theta \cos \phi )^2 \Big |\frac{\varepsilon - 1}{\varepsilon - (\sin \theta \cos \phi )^2} \Big |^2 \cr && \frac{\sin^2 \Big (\frac{d-a}{2} \frac{\omega}{c} (\beta^{-1} - \sin \theta \cos \phi )\Big )}{(1 - \beta \sin \theta \cos \phi )^2} \frac{\sin^2 \Big (N \frac{d}{2} \frac{\omega}{c}\Big (\frac{1}{\beta} - \sin \theta \cos \phi \Big )\Big )}{\sin^2 \Big (\frac{d}{2} \frac{\omega}{c}\Big (\frac{1}{\beta} - \sin \theta \cos \phi \Big )\Big )} e^{-h \frac{2 \omega}{v \gamma} \sqrt{1 + (\beta \gamma \sin \theta \sin \phi )^2}} \times \cr && \Big |(\cos \theta + \sqrt{\varepsilon - \sin^2 \theta})^2 e^{- i b\frac{\omega}{c}\sqrt{\varepsilon - \sin^2 \theta}} - (\cos \theta - \sqrt{\varepsilon - \sin^2 \theta})^2 e^{i b\frac{\omega}{c}\sqrt{\varepsilon - \sin^2 \theta}}\Big |^{-2}\times \cr && \times \Big | (\sqrt{\varepsilon - \sin^2 \theta} - \cos \theta )\Big (\sqrt{1 + (\beta \gamma \sin \theta \sin \phi )^2} - i \beta \gamma \sqrt{\varepsilon - \sin^2{\theta}}\Big ) e^{i b \frac{\omega}{c} \sqrt{\varepsilon - \sin^2{\theta}}} + \cr && + (\sqrt{\varepsilon - \sin^2 \theta} + \cos \theta )\Big (\sqrt{1 + (\beta \gamma \sin \theta \sin \phi )^2} + i \beta \gamma \sqrt{\varepsilon - \sin^2{\theta}}\Big ) e^{-i b \frac{\omega}{c} \sqrt{\varepsilon - \sin^2{\theta}}} - \cr && \quad - 2 \sqrt{\varepsilon - \sin^2 \theta} \Big (i \beta \gamma \cos \theta + \sqrt{1 + (\beta \gamma \sin \theta \sin \phi )^2}\Big ) e^{- b \frac{\omega}{v \gamma} \sqrt{1 + (\beta \gamma \sin \theta \sin \phi )^2}}\Big |^2. \label{Eq48}
\end{eqnarray}

The formula for the energy emitted into the lower half-space (which contains no particle) can be obtained analogously to expressions for the backward radiation intensity in the previous sections. For this
purpose, the charge trajectory must by specularly mapped (relative to the $y$ axis) to the half-space $z<0$. In this case, the field of the particle differs from expression (\ref{Eq45}) in the sign of the $z$-component and in the substitution $z^{\prime} \rightarrow -z^{\prime}$ in the exponent. The magnetic field components of radiation into the half-space $z > 0$ (containing no particle) differ from components (\ref{Eq47}) with the subscript (B) in the reversal of the sign of $b$ in the exponent (as in the previous section). By virtue of these arguments, we obtain the following expressions for the intensity of radiation into the lower half-space:
\begin{eqnarray}
\displaystyle && \frac{d^2W}{d\omega d\Omega}\Big |_{\perp (down)} = \frac{e^2}{\pi^2 c}  \frac{(\beta \gamma )^{-2} \cos^2{\theta}}{1 + (\beta \gamma \sin \theta \sin \phi )^2} \Big |\frac{\varepsilon - 1}{\varepsilon - (\sin \theta \cos \phi )^2} \Big |^2 \frac{\sin^2 \Big (\frac{d-a}{2} \frac{\omega}{c} (\beta^{-1} - \sin \theta \cos \phi )\Big )}{(1 - \beta \sin \theta \cos \phi )^2}\times \cr && \frac{\sin^2 \Big (N \frac{d}{2} \frac{\omega}{c}\Big (\frac{1}{\beta} - \sin \theta \cos \phi \Big )\Big )}{\sin^2 \Big (\frac{d}{2} \frac{\omega}{c}\Big (\frac{1}{\beta} - \sin \theta \cos \phi \Big )\Big )} e^{-h \frac{2 \omega}{v \gamma} \sqrt{1 + (\beta \gamma \sin \theta \sin \phi )^2}} \Big |(\varepsilon \cos \theta + \sqrt{\varepsilon - \sin^2 \theta})^2 e^{- i b\frac{\omega}{c}\sqrt{\varepsilon - \sin^2 \theta}} - \cr && - (\varepsilon \cos \theta - \sqrt{\varepsilon - \sin^2 \theta})^2 e^{i b\frac{\omega}{c}\sqrt{\varepsilon - \sin^2 \theta}}\Big |^{-2}\times \Big | (\sqrt{\varepsilon - \sin^2 \theta} - \varepsilon \cos \theta )\Big (\sqrt{1 + (\beta \gamma \sin \theta \sin \phi )^2} + \cr && + i \beta \gamma \sqrt{\varepsilon - \sin^2{\theta}}\Big ) \Big (i \sin \theta \sqrt{1 + (\beta \gamma \sin \theta \sin \phi )^2} + \sqrt{\varepsilon - \sin^2{\theta}}(\gamma^{-1} \cos \phi + \beta \gamma \sin \theta \sin^2 \phi ) \Big ) \cr && \times e^{i b \frac{\omega}{c} \sqrt{\varepsilon - \sin^2{\theta}}} + (\sqrt{\varepsilon - \sin^2 \theta} + \varepsilon \cos \theta )\Big (\sqrt{1 + (\beta \gamma \sin \theta \sin \phi )^2} - i \beta \gamma \sqrt{\varepsilon - \sin^2{\theta}}\Big ) \times \cr && \Big (i \sin \theta \sqrt{1 + (\beta \gamma \sin \theta \sin \phi )^2} - \sqrt{\varepsilon - \sin^2{\theta}}(\gamma^{-1} \cos \phi + \beta \gamma \sin \theta \sin^2 \phi ) \Big ) e^{-i b \frac{\omega}{c} \sqrt{\varepsilon - \sin^2{\theta}}} - \cr && \quad - 2 \sqrt{\varepsilon - \sin^2 \theta} \Big (i \beta \cos \phi (\varepsilon - \sin^2 \theta) + i \sin \theta (1 + \varepsilon(\beta \gamma \sin \phi )^2) - \varepsilon \cos \theta \times \cr && \qquad \qquad \sqrt{1 + (\beta \gamma \sin \theta \sin \phi )^2} (\gamma^{-1} \cos \phi - \beta \gamma \sin \theta \cos^2 \phi )\Big )
e^{b \frac{\omega}{v \gamma} \sqrt{1 + (\beta \gamma \sin \theta \sin \phi )^2}}\Big |^2, \cr &&
\displaystyle \frac{d^2W}{d\omega d\Omega}\Big |_{\parallel (down)} = \frac{e^2}{\pi^2 c}  \frac{(\beta \gamma )^{-2} \cos^2{\theta}}{1 + (\beta \gamma \sin \theta \sin \phi )^2} \sin^2 \phi (\gamma^{-1} - \beta \gamma \sin \theta \cos \phi )^2 \Big |\frac{\varepsilon - 1}{\varepsilon - (\sin \theta \cos \phi )^2} \Big |^2 \cr && \frac{\sin^2 \Big (\frac{d-a}{2} \frac{\omega}{c} (\beta^{-1} - \sin \theta \cos \phi )\Big )}{(1 - \beta \sin \theta \cos \phi )^2} \frac{\sin^2 \Big (N \frac{d}{2} \frac{\omega}{c}\Big (\frac{1}{\beta} - \sin \theta \cos \phi \Big )\Big )}{\sin^2 \Big (\frac{d}{2} \frac{\omega}{c}\Big (\frac{1}{\beta} - \sin \theta \cos \phi \Big )\Big )} e^{-h \frac{2 \omega}{v \gamma} \sqrt{1 + (\beta \gamma \sin \theta \sin \phi )^2}} \times \cr && \Big |(\cos \theta + \sqrt{\varepsilon - \sin^2 \theta})^2 e^{- i b\frac{\omega}{c}\sqrt{\varepsilon - \sin^2 \theta}} - (\cos \theta - \sqrt{\varepsilon - \sin^2 \theta})^2 e^{i b\frac{\omega}{c}\sqrt{\varepsilon - \sin^2 \theta}}\Big |^{-2}\times \cr && \times \Big | (\sqrt{\varepsilon - \sin^2 \theta} - \cos \theta )\Big (\sqrt{1 + (\beta \gamma \sin \theta \sin \phi )^2} + i \beta \gamma \sqrt{\varepsilon - \sin^2{\theta}}\Big ) e^{i b \frac{\omega}{c} \sqrt{\varepsilon - \sin^2{\theta}}} + \cr && + (\sqrt{\varepsilon - \sin^2 \theta} + \cos \theta )\Big (\sqrt{1 + (\beta \gamma \sin \theta \sin \phi )^2} - i \beta \gamma \sqrt{\varepsilon - \sin^2{\theta}}\Big ) e^{-i b \frac{\omega}{c} \sqrt{\varepsilon - \sin^2{\theta}}} - \cr && \quad - 2 \sqrt{\varepsilon - \sin^2 \theta} \Big (-i \beta \gamma \cos \theta + \sqrt{1 + (\beta \gamma \sin \theta \sin \phi )^2}\Big ) e^{b \frac{\omega}{v \gamma} \sqrt{1 + (\beta \gamma \sin \theta \sin \phi )^2}}\Big |^2. \label{Eq49}
\end{eqnarray}
Comparison of formulas (\ref{Eq48}) and (\ref{Eq49}) shows that the transition from radiation "above the grating" to that "under the grating" is performed by the following substitution analogous to (\ref{Eq32}) (everywhere except in the exponents):
\begin{eqnarray}
\displaystyle && \sqrt{\varepsilon - \sin^2 \theta} \rightarrow - \sqrt{\varepsilon - \sin^2 \theta}, \ \cos{\theta}\rightarrow - \cos {\theta}, \cr && e^{-b \frac{\omega}{v \gamma} \sqrt{1 + (\beta \gamma \sin \theta \sin \phi )^2}}\rightarrow e^{b \frac{\omega}{v \gamma} \sqrt{1 + (\beta \gamma \sin \theta \sin \phi )^2}}. \label{Eq50}
\end{eqnarray} 
It should be emphasized that impact parameter $h$ appearing in formula (\ref{Eq49}) is measured, as in expressions (\ref{Eq48}), from the plane $z=0$; however, in the "under-the-grating" radiation geometry, the shortest distance between the charge trajectory and the grating is $h - b$ and not $h$. The reason for quantity $h$ (and not $h-b$) appearing in formula (\ref{Eq49}) is the fact that we consider radiation into the half-space $z>0$, and in the limit of a conducting grating, the skin effect appears in the vicinity of $z=0$ plane.

Formulas (\ref{Eq48}) and (\ref{Eq49}) for the radiation intensity can also be written in variables $\vartheta, \varphi$, which are conventionally used in the geometry of Smith-Purcell radiation. The relations between these variables and those used in this study have the form
\begin{eqnarray}
\displaystyle && \sin \theta \sin \phi = \sin \vartheta \sin \varphi, \ \sin \theta \cos \phi = \cos \vartheta, \ \cos \theta = \sin \vartheta \cos \varphi, \label{Eq51}
\end{eqnarray}
which leads, in particular, to the following relations:
\begin{eqnarray}
\displaystyle \sin^2 \theta = 1 - \sin^2 \vartheta \cos^2 \varphi, \ \sin \phi = \frac{\sin \vartheta \sin \varphi}{\sqrt{1 - (\sin \vartheta \cos \varphi )^2}}, \ \cos \phi = \frac{\cos \vartheta}{\sqrt{1 - (\sin \vartheta \cos \varphi )^2}}. \label{Eq52}
\end{eqnarray}

The expressions for the emitted energy contain the factor
\begin{eqnarray}
\displaystyle && F_{strip} = 4 \sin^2 \Big (\frac{d-a}{2} \frac{\omega}{v} (1 - \beta \cos \vartheta )\Big ), \label{Eq53}
\end{eqnarray}
describing the interference of the waves emitted on the left and right edges of a strip of length $d-a$. If the
charge moves only near a single long strip, this term divided by $(1 - \beta \cos \vartheta )^2$ gives, analogously to (\ref{Eq19}), the delta-function determining the condition for Cherenkov radiation in a vacuum. However, the argument of the delta-function $1 - \beta \cos \vartheta $ never vanishes, which corresponds to the condition for the total internal reflection of radiation from the upper and lower faces of the strip. Waves of Cherenkov radiation emerge into the vacuum only from the side facets of the grating at angles $\vartheta \precsim b/d$, for which the model considered here is no longer applicable.

For a large number of periods $N \gg 1$, the formulas for the emitted energy contain another delta-function
\begin{eqnarray}
\displaystyle && F_N = \frac{\sin^2{\Big (N \frac{d}{2}\frac{\omega}{c}(\frac{1}{\beta} - \cos \vartheta)\Big )}}{\sin^2{\Big (\frac{d}{2}\frac{\omega}{c}(\frac{1}{\beta} - \cos \vartheta)\Big )}} \rightarrow 2 \pi N \sum \limits_{m = 1}^{\infty}\delta \Big (d \frac{\omega}{c}\Big (\frac{1}{\beta} - \cos \vartheta \Big ) - 2 \pi m \Big ), \label{Eq54}
\end{eqnarray}
whose zeros give the well-known Smith-Purcell relation
\begin{eqnarray}
\displaystyle && \lambda_m = \frac{d}{m} \Big (\frac{1}{\beta} - \cos \vartheta \Big ). \label{Eq55}
\end{eqnarray}
It should be emphasized that formulas (\ref{Eq48}) and (\ref{Eq49}) derived above permit factorization of the type
\begin{eqnarray}
\displaystyle && \frac{d^2W}{d\omega d\Omega} = \frac{d^2W}{d\omega d\Omega}\Big |_{screen} F_{strip} F_N, \label{Eq55a}
\end{eqnarray}
where $d^2W/(d\omega d\Omega)_{screen}$ is the intensity of radiation generated when a charge moves over a semi-infinite rectangular screen parallel to the plane through which we observe the emergence of radiation into the vacuum. This circumstance is also a consequence of the condition $a \ll \lambda$ used here. In the models of Smith-Purcell radiation based on exact formulation of the boundary conditions in the gaps between perfectly conducting strips, as well as in the models for gratings with a periodic surface profile, the above factorization is not observed (see, for example, \cite{Sh, Kube, van}).

In the limit of a perfectly conducting grating ($\varepsilon^{\prime \prime}\rightarrow \infty$) the energy emitted into both half-spaces is the same and is given by
\begin{eqnarray}
&& \displaystyle \frac{d^2 W}{d \omega d \Omega} = \frac{d^2 W}{d \omega d \Omega}\Big |_\perp + \frac{d^2 W}{d \omega d \Omega}\Big |_\parallel = \cr && \qquad = \frac{e^2}{\pi^2 c} \frac{\gamma^{-2} (1 - (\sin \vartheta \sin \varphi )^2 ) + 2 \beta \cos \vartheta (\sin \vartheta \sin \varphi )^2
+ (\beta \gamma )^2 \sin^4 \vartheta \sin^2 \varphi}{1 + (\beta \gamma \sin \vartheta \sin \varphi)^2} \cr && \qquad \qquad \displaystyle
\times \frac{\sin^2{\Big (\frac{d - a}{2} \frac{\omega}{c}(\frac{1}{\beta} - \cos \vartheta)\Big )}}{(1 - \beta \cos \vartheta)^2}
\frac{\sin^2{\Big (N \frac{d}{2}\frac{\omega}{c}(\frac{1}{\beta} - \cos \vartheta)\Big )}}{\sin^2{\Big (\frac{d}{2}\frac{\omega}{c}(\frac{1}{\beta} - \cos \vartheta)\Big )}}
e^{-h \frac{2\omega}{v \gamma}\sqrt{1 + (\beta \gamma \sin \vartheta \sin \varphi)^2}}. \label{Eq56}
\end{eqnarray}
As expected, the dependence on the thickness of the grating has disappeared (skin effect). For a large number of periods, this expression can be integrated with respect to frequency using formula (\ref{Eq54}). In this case, the angular distribution of the energy emitted at the $m$th harmonic has a simple form:
\begin{eqnarray}
&& \displaystyle \frac{1}{N}\frac{d W_m}{d \Omega} = \frac{2 e^2 \beta}{\pi d} R(\gamma, \vartheta, \varphi ) 
\frac{\sin^2{\Big (\pi m \frac{d - a}{d}\Big )}}{(1 - \beta \cos \vartheta)^3} e^{-h \frac{4 \pi}{\beta \gamma \lambda_m}\sqrt{1 + (\beta \gamma \sin \vartheta \sin \varphi)^2}}, \label{Eq57}
\end{eqnarray}
where $R(\gamma, \vartheta, \varphi )$ denotes the ``angular'' part of formula (\ref{Eq56}):
\begin{eqnarray}
& \displaystyle R(\gamma, \vartheta, \varphi ) = \frac{\gamma^{-2} (1 - (\sin \vartheta \sin \varphi )^2 ) + 2 \beta \cos \vartheta (\sin \vartheta \sin \varphi )^2
+ (\beta \gamma )^2 \sin^4 \vartheta \sin^2 \varphi}{1 + (\beta \gamma \sin \vartheta \sin \varphi)^2}. \label{EqR}
\end{eqnarray}
It can easily be seen that in the ultrarelativistic limit ($\gamma \gg 1$) the following relation holds for non-zero angles $\vartheta$ and $\varphi$:
\begin{eqnarray}
&& \displaystyle R(\gamma, \vartheta, \varphi ) \rightarrow \sin^2 \vartheta. \label{Eq58}
\end{eqnarray}
Formula (\ref{Eq57}) in this case completely coincides with the expression derived in the well-known model of surface currents presented in Ref.\cite{Br}.

It is significant that formula (\ref{Eq56}) for a perfectly conducting infinitely thin grating can also be derived using a completely different approach. Namely, we can assume that radiation is generated by the surface current induced on the grating. Such an approach, which was developed in \cite{Br} in the problem of Smith-Purcell radiation (see also \cite{Kes}), has the following disadvantage: the surface current is usually assumed to have only two tangential components (analogously to the theory of diffraction of ordinary plane waves). It was shown in \cite{PLA} that in the general case, when the field incident on a perfectly conducting plane has all three components, the surface density of the induced current must also have three components including the component normal to the surface. The normal component of the current density may be disregarded only in the ultrarelativistic case, when field $\bold E^0$ of the particle can be treated as transverse. The intensity of Smith-Purcell radiation determined using such a model of the \textit{generalized} surface current completely coincides with expression (\ref{Eq56}) \cite{PHD}. As expected, in the ultrarelativistic limit, this result coincides with that calculated in \cite{Br} in the model with a surface current with zero normal component. This probably explains the conformity of predictions of the model \cite{Br} to the experimental results \cite{Doucas} obtained for an electron bunch with an energy of 28.5 GeV.

It should also be noted that analogously to the previous section, the presence of factor $1 + (\beta \gamma \sin \vartheta \sin \varphi)^2$ in the denominators of formulas (\ref{Eq48}), (\ref{Eq49}), and (\ref{Eq56}) leads to concentration of radiation emitted at large angles $\vartheta \gg \gamma^{-1}$ in the plane perpendicular to the plane of the grating (i.e., close to small azimuthal angles $\varphi \precsim \gamma^{-1}$). The minimum observed strictly at angle $\varphi = 0$ disappears in the ultrarelativistic limit in accordance with expression (\ref{Eq58}).

\section{\large{Discussion}}
\label{Sect4}

All types of radiation considered in the present paper (Cherenkov radiation, transition radiation, diffraction radiation, and Smith-Purcell radiation) physically originate as a result of dynamic polarization
of atoms and molecules of the medium by the field of an external source. Therefore, it is natural to use a unified formalism in their description. We have demonstrated that the method of induced currents makes it possible to determine the characteristics of various types of polarization radiation and is applicable for a wide class of shapes of the surfaces. In this study, the actual dielectric properties of a substance are systematically taken into account in problems of polarization radiation in media with sharp boundaries of an intricate shape. In the limiting case of perfect conductivity, all results coincide with those available in the literature. It may be shown that in another limiting case of small permittivity ($|\varepsilon (\omega) - 1| \ll 1$) all the solutions obtained also coincide with corresponding results of perturbation theory (see e.g. Refs.\cite{T-PLA, T-PRE, Mono, Sysh}).

The solutions obtained in this study can be used in various applications of polarization radiation, in particular, in the development of new types of monochromatic radiation sources and in the physics of accelerators. For each of the problems considered in this paper, the applicability boundaries of the results can be indicated.

1. In the problem of radiation in a cylindrical channel considered in Section 2, we assumed that the screen thickness $d$ is so small that radiation emerging through the outer and inner walls of the cylinder can be ignored. In the relativistic case, the condition for the applicability of the solution to this problem can be written in the form of a single inequality. Indeed, the effective region of the screen participating in the formation of radiation is $(b-a)_{eff} \approx \gamma \lambda$. Therefore, the waves emitted through the outer surface of the cylinder may play a significant role only if the outer radius $b \ll \gamma \lambda$. On the other hand, only Cherenkov radiation waves are in fact incident on the outer surface of the cylinder (when the relevant condition is satisfied) because diffraction radiation waves propagate at small angles $\theta \sim \gamma^{-1} \ll 1$ in the relativistic case. Cherenkov radiation incident on the outer as well as inner surface of the cylinder experiences total internal reflection and can emerge into the vacuum through the front and rear faces of the screens at angles $\theta \sim \pi/2$. If we confine our analysis to the angles of emission not
too close to $\pi/2$, the condition for the applicability of our results for all values of channel radius $a \gg \lambda$ and screen width $(b-a) \succsim \gamma \lambda$ has the following simple form:
\begin{eqnarray}
&& \displaystyle d \ll \gamma \lambda. \label{Eq59}
\end{eqnarray}
For experimental values of $d$ larger than a few millimeters, it follows that in the optical and infrared ranges, the solution obtained here can be used only for ultrarelativistic particles. For frequencies from the X-ray range, we can disregard re-reflections and constraint (\ref{Eq59}) imposed on the screen thickness becomes invalid.

2. In the problem of radiation from a rectangular screen in Section 3, we must consider two limiting values of the screen length: $a > \gamma \lambda$ and $a \precsim \gamma \lambda$. 
In the former case, the screen length can be assumed to be infinitely large, and the condition for the admissible screen thickness coincides with (\ref{Eq59}). 
In the latter case, the geometrical size of the screen is important; therefore, the following condition must be satisfied:
\begin{eqnarray}
&& \displaystyle d \ll a. \label{Eq60}
\end{eqnarray}

3. The condition for the applicability of the solution to the problem of radiation from the grating in Section 4 (inequalities (\ref{Eq42b}), (\ref{Eq43}), and (\ref{Eq44})) is even
more important because these conditions also determine the applicability boundaries for the solutions existing earlier and obtained for perfectly conducting gratings using the models of surface currents, for
which these boundaries are not obvious. Indeed, the field of radiation from a grating in vacuum in these models can be presented in the form of the vector sum of the field of radiation from each strip. The expression for the density of the current induced on the strip surface by the field of a moving charge is sought in different models using different methods (see, for example, \cite{Br, PHD, Pap, Kes}). For a grating consisting of strips with a finite permittivity and separated by vacuum gaps, such a summation can be carried out only if the strips are closely spaced ($a \ll \lambda \sim d$) and multiple re-reflection of waves in the gaps can be ignored. Naturally, this condition remains in force in the case of perfectly conducting infinitely thin strips also; however, it is not specified in models of surface currents, in which the problem is initially solved for a perfect conductor.

Indeed, it appears at first glance that re-reflections can be disregarded for infinitely thin strips, and the results (formula (\ref{Eq56})) must be applicable for any width
$a$ of vacuum gaps. However, as emphasized above, on account of the real dielectric properties of the grating material, the vector summation of the radiation fields from each strip is carried out for the interior of the medium and not for the vacuum. Such a procedure has sense only if the strips are closely spaced. The application of Fresnel coefficients as well as formula (\ref{Eq24}) for the radiation intensity, which were derived for an infinitely large plate, is also limited by condition (\ref{Eq43}) in the problem of the grating. In the models of surface currents, the expression for the current itself is usually derived by disregarding distortion at the edges of a strip (i.e., as for a perfectly conducting semi-plane \cite{Br, Pap}). This approximation precisely corresponds to disregarding re-reflections of radiation between the strips as well as the use of the Fresnel coefficients for an infinitely large plate in the problem with a finite conductivity of the grating. Since the gratings with vacuum gaps between the strips are conventionally used for generating radiation in the terahertz and microwave ranges, the inequality $a \ll d$ can easily be realized in experiments.

In conclusion, we note that the well-known fact of the zero radiation intensity at even harmonics for $a = d/2$, which is predicted by the models of surface currents (see formula (\ref{Eq57}) as well as \cite{Br, Pap}) and which is absent in the van den Berg's model \cite{van}, can be explained by the fact that the models of surface currents for the given vacuum gap width
can be used only for qualitative analysis.

\textbf{Acknowledgments.} The author wishes to thank A.P. Potylitsyn, A.A. Tishchenko, and L.G. Sukhikh for constructive critical remarks and fruitful discussions. This work is partially supported by the Ministry of Education and Science of the Russian Federation under the program ``Scientific and Scientific-Pedagogical Personnel of Innovative Russia for 2009-2013,'' contracts $\Pi$1199 (June 4, 2010), $\Pi$1145 (August 27, 2009), and 02.740.11.0238.

\begin{figure*}
\center \includegraphics[width=12.00cm, height=7.00cm]{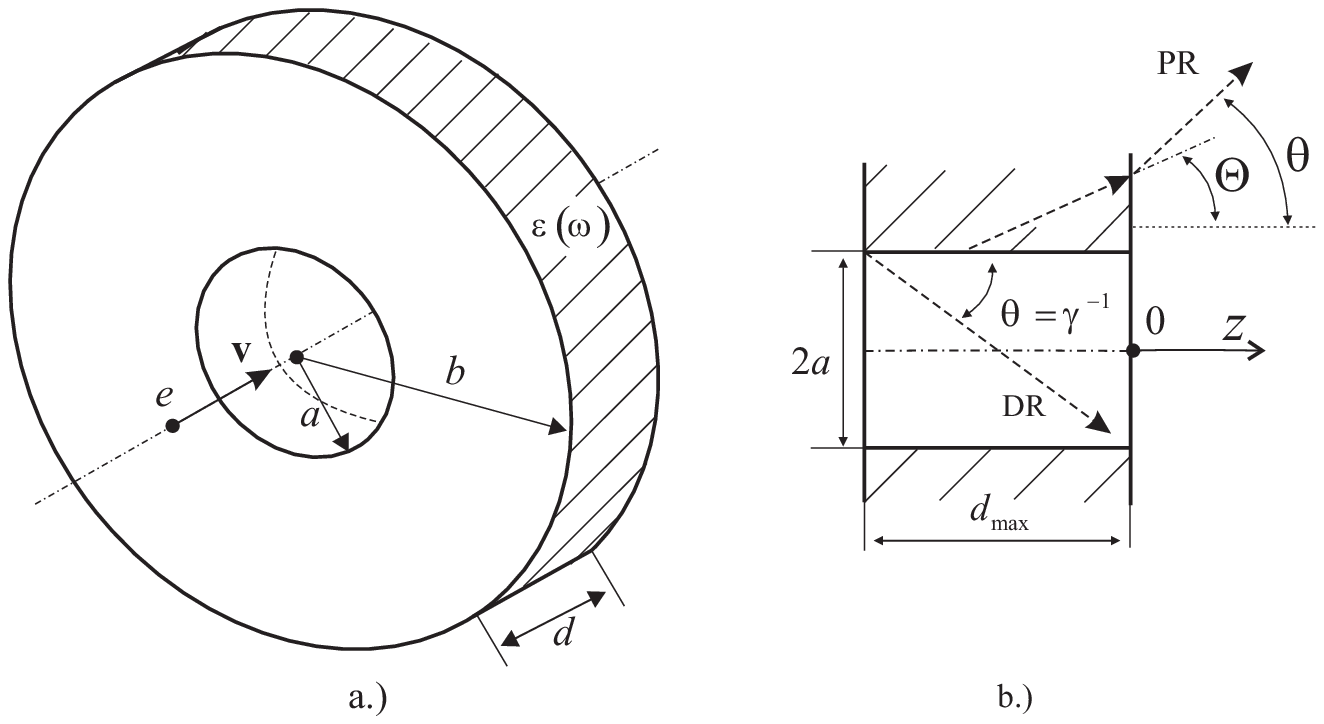}
\caption{\label{Fig1} (a) Diagram of generation of polarization radiation (PR) by a charged particle moving in a cylindrical channel. (b) Determination of the applicability boundaries of the model; DR is diffraction radiation.}
\end{figure*}
\begin{figure*}
\center \includegraphics[width=5.00cm, height=6.50cm]{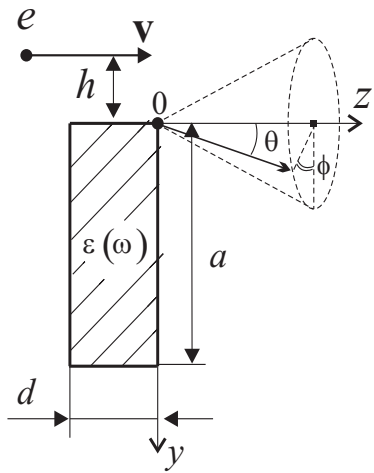}
\caption{\label{Fig2} Diagram of generation of polarization radiation by a charged particle moving near a rectangular screen.}
\end{figure*}
\begin{figure*}
\center \includegraphics[width=9.00cm, height=6.00cm]{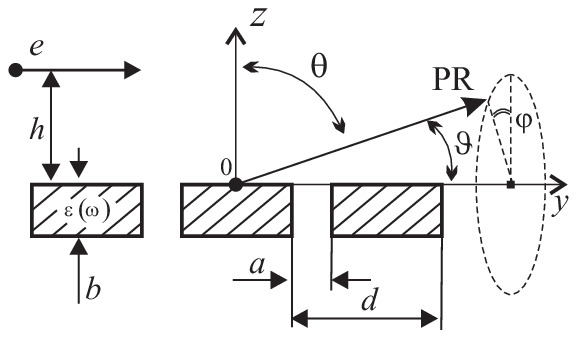}
\caption{\label{Fig3} Diagram of generation of polarization radiation (PR) by a charged particle moving near a grating.}
\end{figure*}


\begin{thebibliography}{35}

\bibitem{A}
M.Ya. Amusia, Rad. Phys. Chem. {\bf 75}, 1232 (2006).

\bibitem{B}
V.E. Pafomov, Proc. P.N. Lebedev Phys. Inst. \textbf{44}, 25 (1971).

\bibitem{J}
D.V. Karlovets and A.P. Potylitsyn, JETP \textbf{107} (5), 755 (2008).

\bibitem{PLA}
D.V. Karlovets and A.P. Potylitsyn, Phys. Lett. A \textbf{373}, 1988 (2009).

\bibitem{VDB}
P.M. van den Berg, A.J.A. Nicia, J. Phys. A: Math. Gen. {\bf 9}, 1133 (1976).

\bibitem{G}
I. A. Gilinskii, Electromagnetic Surface Phenomena (Nauka, Novosibirsk, 1990) [in Russian].

\bibitem{Sh} 
V.P. Shestopalov, The Smith-Purcell effect, Nova Science Publ., N.-Y. (1998).

\bibitem{Kube} 
G. Kube, Nucl. Instrum. Meth. B {\bf 227}, 180 (2005).

\bibitem{Mono}
A.P. Potylitsyn, M.I. Ryazanov, M.N. Strikhanov and A.A. Tishchenko, Diffraction radiation from relativistic particles, Springer, Berlin (2010).

\bibitem{D}
L. Durand, Phys. Rev. D {\bf 11}, 89 (1975).

\bibitem{T-PRE}
A.A. Tishchenko, A.P. Potylitsyn, M.N. Strikhanov, Phys. Rev. E {\bf 70}, 066501 (2004).

\bibitem{T-PLA}
A.A. Tishchenko, A.P. Potylitsyn, M.N. Strikhanov, Phys. Lett. A {\bf 359}, 509 (2006).

\bibitem{Sysh}
N.F. Shul'ga, V.V. Syshchenko, J. Phys.: Conf. Ser. {\bf 236}, 012010 (2010).

\bibitem{Ryaz-RF}
M. I. Ryazanov, JETP \textbf{100} (3), 468 (2005).

\bibitem{Topt}
I. N. Toptygin, Modern Electrodynamics, Vol. 2: Theory of Electromagnetic Phenomena in Matter (Regulyarnaya i Khaoticheskaya Dinamika, Moscow, 2005) [in Russian].

\bibitem{Bol-61}
B. M. Bolotovskii, Sov. Phys.-Usp. \textbf{4}, 781 (1961).

\bibitem{Ryzh}
I.S. Gradshteyn and I.M. Ryzhik, Table of Integrals, Series, and Products (Fizmatlit, Moscow, 1963; Academic, New York, 1994).

\bibitem{Z}
V.P. Zrelov, M. Klimanova, V.P. Lupiltsev et al., Nucl. Instrum. Meth. {\bf 215}, 141 (1983).

\bibitem{Gr}
M.I. Ryazanov, Electrodynamics of Condensed Matter (Nauka, Moscow, 1984) [in Russian].

\bibitem{L}
L.D. Landau and E.M. Lifshitz, Course of Theoretical Physics, Vol. 8: Electrodynamics of Continuous Media (Butterworth-Heinemann, Oxford, 2004; Fizmatlit, Moscow, 2005).

\bibitem{JL}
D.V. Karlovets and A.P. Potylitsyn, JETP Lett. \textbf{90} (5), 326 (2009).

\bibitem{G-T}
V.L. Ginzburg and V.N. Tsytovich, Transition Radiation and Transition Scattering (Nauka, Moscow, 1984; Adam Hilger, New York, 1990).

\bibitem{Term}
M. L. Ter-Mikaelyan, High-Energy Electromagnetic Processes in Condensed Media (Academy of Science of the ArmSSR, Yerevan, 1969; Wiley, New York, 1972).

\bibitem{Gar}
G. M. Garibyan and Yan Shi, X-Ray Transition Radiation (Academy of Science of the ArmSSR, Yerevan,1983) [in Russian].

\bibitem{Dn}
Yu.N. Dnestrovskii and D.P. Kostomarov, Sov. Phys. Dokl. \textbf{4}, 158 (1959).

\bibitem{Bol}
B.M. Bolotovskii and E.A. Galst'yan, Phys.-Usp. \textbf{43} (8), 755 (2000).

\bibitem{X}
D. Xiang, W.-H. Huang, Y.-Z. Lin, S.-J. Park, and I. S. Ko, Phys. Rev. ST Accel. Beams {\bf 11}, 024001 (2008).

\bibitem{van} 
P.M. van den Berg, J. Opt. Soc. Am. {\bf 63}, 1588 (1973).

\bibitem{Br}
J.H. Brownell, J. Walsh, G. Doucas, Phys. Rev. E {\bf 57}, 1075 (1998).

\bibitem{Kes} A.S. Kesar, Phys. Rev. ST Accel. Beams {\bf 8}, 072801 (2005).

\bibitem{PHD}
D.V. Karlovets, Candidate's Dissertation in Mathematical Physics (Tomsk Polytechnical University, Tomsk, 2008) [in Russian].

\bibitem{Doucas}
V. Blackmore, G. Doucas, C. Perry et al., Phys. Rev. ST Accel. Beams {\bf 12}, 032803 (2009).

\bibitem{Pap}
A.P. Potylitsyn, Phys. Lett. A {\bf 238}, 112 (1998).



\end{thebibliography}
\end{document}